\documentclass[12pt]{article}
\usepackage{graphicx}
\usepackage{amsfonts,amssymb}
\usepackage{citehack,cite}

\newcommand{\no}{\nonumber}
\newcommand{\non}{\nonumber \\}
\newcommand{\ve}[1]{{\bf #1}}
\newcommand{\be}{\begin{equation}}
\newcommand{\ee}{\end{equation}}
\newcommand{\bea}{\begin{eqnarray}}
\newcommand{\eea}{\end{eqnarray}}
\newcommand{\sli}{\sum\limits}

\newcommand{\lp}{\left (}
\newcommand{\rp}{\right )}
\newcommand{\lb}{\left [}
\newcommand{\rb}{\right ]}

\newcommand{\ld}{\left .}
\newcommand{\rd}{\right .}
\newcommand{\vk}{\ve{k}}
\newcommand{\vl}{\ve{l}}
\newcommand{\cB}{{\cal B}}
\newcommand{\rhok}{\rho_{\ve{k}}}
\newcommand{\rhomk}{\rho_{-\ve{k}}}
\newcommand{\etak}{\eta_{\ve{k}}}

\begin{document}

\begin{center}
{\bf FLUID CRITICAL BEHAVIOR AT LIQUID--GAS PHASE TRANSITION:
ANALYTIC METHOD FOR MICROSCOPIC DESCRIPTION}
\end{center}

\begin{center}
{\sc I.V. Pylyuk}
\end{center}

\begin{center}
{\it Institute for Condensed Matter Physics  \\
of the National Academy of Sciences of Ukraine, \\
1~Svientsitskii Str., 79011 Lviv, Ukraine} \\
E-mail: piv@icmp.lviv.ua
\end{center}

\vspace{0.5cm}

{\small
The behavior of fluids in the vicinity of the liquid--gas critical
point is studied within the cell fluid model framework.
The analytic method for deriving the equation of state of a cell
fluid model in the low-temperature region ($T<T_c$) is
developed using the renormalization group transformation within
the collective variables approach. Mathematical description within
the grand canonical ensemble is illustrated by an example of the Morse
interaction potential possessing the Fourier transform.
A specific feature of the proposed method lies in the possibility
to use exclusively microscopic characteristics of a fluid
(parameters of the interaction potential) for calculating macroscopic
quantities (pressure and other thermodynamic quantities) without
involving the hard-spheres reference system.
The grand partition function, thermodynamic potential, and
equation of state of the model near the critical point are derived
taking into account the non-Gaussian (quartic) distribution of order
parameter fluctuations. A nonlinear equation, which links the density
and the chemical potential, is presented and solved.
Graphs of the dependence of the density on the chemical potential
are plotted for various values of the relative temperature.
The numerical estimates of the critical-point parameters for potassium,
obtained in addition to the estimates for sodium, are given.
The calculated critical-point parameters for liquid alkali metals
(sodium and potassium) are in accord with experimental data.
The coexistence curve for sodium is plotted and compared with other
authors' data in the immediate vicinity of $T_c$, where theoretical and
experimental researches are difficult to carry out. The differences
between the obtained results and the earlier published results for
$T>T_c$ are discussed.
}

\vspace{0.5cm}


\section*{List of notations}
\label{sec:0}

\begin{tabular}{p{2.00cm}p{10.24cm}}
\multicolumn{1}{l}{\emph{Symbol}} & \multicolumn{1}{c}{\emph{Meaning}} \\[3pt]
$N_v$ & number of cells \\
$c$ & linear cell size \\
$v$ & volume of cubic cell \\
$T$ & temperature \\
$T_c$ & critical temperature \\
$\Xi$ & grand partition function \\
$W(k)$ & Fourier transform of effective potential \\
$Q_n$ & partial partition function of $n$th layer \\
$\Xi_{IGR}$ & contribution to grand partition function from inverse
Gaussian regime \\
$\tilde h$ & renormalized external field \\
$h_{cm}$ & temperature field \\
$\tau$ & relative temperature \\
$\Omega$ & thermodynamic potential \\
$\Omega_\mu$ & thermodynamic potential contribution corresponding to partition
function component $G_\mu$ \\
$\Omega_r$ & thermodynamic potential contribution corresponding to partition
function component $(Q(r_0))^{N_v}$ \\
$\Omega^{(-)}_{CR}$ & contribution to thermodynamic potential from critical
regime, $T<T_c$ \\
$\gamma_{01}, \gamma_{02}, \gamma_{03}$ & coefficients for analytic
part of $\Omega^{(-)}_{CR}$ \\
$\Omega_{CR}^{(s)'}$ & singular part of $\Omega^{(-)}_{CR}$ \\
$\bar\gamma^{(-)}$ & coefficient for singular part $\Omega_{CR}^{(s)'}$, $T<T_c$ \\
$\Omega_{TR}^{(-)}$ & contribution to thermodynamic potential from transition
region, $T<T_c$ \\
$\Omega_{IGR}$ & contribution to thermodynamic potential from inverse
Gaussian regime \\
$\Omega_a$ & analytic part of thermodynamic potential \\
$\Omega_s^{(-)}$ & nonanalytic part of thermodynamic potential, $T<T_c$ \\
$\Omega_0^{(-)}$ & thermodynamic potential part associated with shift of
variable $\rho_0$, $T<T_c$ \\
$\bar n$ & average density \\
$\mu, M$ & chemical potentials \\
$P$ & pressure \\
\end{tabular}

\section{Introduction}
\label{sec:1}

For more than a century, the behavior of multiparticle systems in the liquid
and gas phases attracts the attention of scientists. The task of the theoretical
description of the properties of fluids on the level of microscopic interactions
between particles remains relevant \cite{hm113}. Especially significant
is the problem of describing fluid in the vicinity of and below the critical
temperature $T_c$. Below $T_c$, two phases can coexist (a gas at low density and
a liquid at high density).

The study of the behavior of fluids is often carried out with the use of the concept
reference system. The role of the latter is usually played by the system of
hard spheres (see, for example, \cite{km189,bs104,r110,y114}).
The novelty of the approach developed in this paper is to use
exclusively microscopic characteristics of the model (parameters of the interaction
potential) for obtaining macroscopic quantities (pressure and other thermodynamic
quantities) without involving the hard-spheres system. The temperature region
below the critical temperature $T_c$ is considered. The refusal to use
the hard-spheres reference system within the collective variables approach
\cite{ymo287,YuKP_2001} is due to several reasons.
One of them (in the presence of the hard-spheres reference
system) is the use of different spaces of variables in calculating the Jacobian of transition
to collective variables and grand partition function. If the Jacobian of the transition
is calculated in Cartesian space, then the grand partition function of the fluid system
is calculated in the space of the collective variables. Methods of calculation in the two
spaces mentioned above are fundamentally different, and therefore require the use
of different approximations that are difficult to reconcile with each other
(to determine common accuracy). The problem of coordinating accuracy in different spaces
as well as the problem of reconciling the average number of particles in the reference system
and their total number can be avoided in the mathematical description without involving
the hard-spheres reference system. In the framework of this approach, the reference system
is not introduced from the outside, but appears at a certain stage in the calculation of
the grand partition function and is not related to the presence of additional chemical
potential. Here one chemical potential for fluid is used.

The theoretical description of the critical behavior of a simple fluid is carried out within
the framework of a cell model. The volume of the system $V$ composed of $N$ interacting
particles is conventionally divided into $N_v$ cells, each of volume $v=V/N_v=c^3$ ($c$ is
the linear cell size). It should be noted that, in contrast to a cell gas model (where it
is assumed that a cell may contain only one particle or does not contain any particle)
\cite{rebenko_13,rebenko_15}, a cell within this approach
may contain more than one particle. Instead of the distance between
the particles, the distance between the centers of the cells is introduced. The cell
interaction potential is chosen in the form of the Morse potential possessing the Fourier
transform. Despite the great successes in the investigation of Morse fluids
made by means of various methods (for example, the $NpT$ plus test particle method
\cite{okumura_00}, the grand-canonical transition matrix Monte Carlo
method \cite{singh}, the integral equations approach \cite{apf_11},
molecular dynamics simulations in a canonical ensemble \cite{martinez}),
the study and statistical description of the behavior of the mentioned fluids near
the critical point on the microscopic level without any general assumptions
are still of interest.

The present theoretical studies supplement the results of paper \cite{kpd118}, in which
a method for calculating the equation of state of the fluid system is developed
in the case of $T>T_c$. Parameters of interaction potential, which are necessary for
quantitative estimates, correspond to data for Na (sodium) and K (potassiun)
\cite{singh}. Expressions for the quantities mentioned and
not given here can be found in \cite{kpd118}.
It should be noted that the temperature region $T<T_c$ differs significantly
from the region $T>T_c$. The reason is that in the system at $T<T_c$ the order
parameter appears. The influence of this order parameter will be taken into account
in the process of determining the system's point of exit from
the critical fluctuation regime.

Briefly describe the structure of the present paper. Sec.~\ref{sec:2}, which follows
the introduction (Sec.~\ref{sec:1}), contains the basic output relations, which are valid for
temperatures below $T_c$. Here the expression for the grand partition function of the model
that it is necessary to calculate for obtaining the thermodynamic potential as well as
formulas for the quantities included in this expression are presented. Scheme for
calculation of the thermodynamic potential is given in Sec.~\ref{sec:3}.
The basic idea of such a calculation within the approach of collective variables
lies in the separate inclusion of contributions from short-wave and long-wave
fluctuations of the order parameter (see, for example, \cite{ykp202}).
Short-wave fluctuations are characterized by
a renormalization group symmetry and are described by a non-Gaussian
distribution. They correspond to the region of the critical regime.
The region of the inverse Gaussian regime is associated with long-wave fluctuations.
The calculation of the contribution to the thermodynamic potential from
long-wave fluctuations is based on the use of the Gaussian distribution as
the basis one. Here, we have developed a direct method of calculations
with the results obtained by taking into account the short-wave fluctuations
as initial parameters. In the same Sec.~\ref{sec:3}, the expression for
short-wave part of the thermodynamic potential (the contribution from the critical regime
of fluctuations) is presented. The procedure for calculating the long-wave part of the thermodynamic
potential (the contribution from the inverse Gaussian regime of fluctuations) is developed
in Sec.~\ref{sec:4}. Using the obtained results, the complete expression for the thermodynamic
potential is written. A method for constructing the equation of state of a cell fluid model
for $T<T_c$ is described in Sec.~\ref{sec:5}, taking into account non-Gaussian fluctuations.
The critical-point parameters calculated for Na and K as well as the binodal
curve obtained for Na in the immediate vicinity of $T_c$ are given.
Concise conclusions are presented in Sec.~\ref{sec:6}. Some supplementary interim
results are provided in two appendices. This makes it easier to read the article.

\section{Basic expressions at temperatures below $T_c$}
\label{sec:2}

The grand partition function of the cell fluid model can be written in the following
form \cite{kpd118,kdp117}:
\be
\Xi = \sli_{N=0}^{\infty} \frac{(z)^N}{N!} \int \limits_{V} (dx)^N
\exp \left[-\frac{\beta}{2} \sli_{\vl_1,\vl_2\in\Lambda}
\tilde U_{l_{12}} \rho_{\vl_1} (\eta) \rho_{\vl_2} (\eta) \right].
\label{0d1fb}
\ee
Here $z = e^{\beta \mu}$ is the activity, $\beta=1/(kT)$
is the inverse temperature, and $\mu$ is the chemical potential.
Integration with respect to coordinates of all the particles
$x_i = (x_{i}^{(1)},x_{i}^{(2)}, x_{i}^{(3)})$ is noted as
$\int \limits_{V} (dx)^N = \int \limits_{V} dx_1 \cdots \int \limits_{V} dx_N$,
and $\eta = \{ x_1 , \ldots , x_N \}$ is the set of coordinates.
The interaction potential $\tilde U_{l_{12}}$ is a function of the distance
$l_{12}= |\ve{l}_{1}-\ve{l}_{2}|$ between cells. Each vector $\ve{l}_i$ belongs
to the set
\[
\Lambda =\Big\{ \vl = (l_1, l_2, l_3)|l_i = c m_i;
 m_i=1,2,\ldots,N_a; i=1,2,3; N_v=N_a^3 \Big\},
\]
where $N_a$ is the number of cells along each axis. The occupation numbers of cells
\be
\rho_{\vl}(\eta) = \sli_{x \in \eta} I_{\Delta_{\vl}(x)}
\label{0d2fb}
\ee
appearing in Eq. (\ref{0d1fb}) are defined by
the characteristic functions (indicators)
\be
I_{\Delta_{\vl}(x)} = \left\{
\begin{array}{l}
1,\quad \texttt{if} \quad x\in\Delta_{\vl} \\
0,\quad \texttt{if} \quad x\notin\Delta_{\vl},
\end{array} \rd
\label{0d3fb}
\ee
which identify the particles in each cubic cell
$\Delta_{\vl} = (-c/2,c/2]^3 \subset \mathbb{R}^3$ and their contribution to
the interaction of a model. In further calculations, the interaction
potential $\tilde U_{l_{12}}$ will be chosen in the form of the Morse potential:
\be
\tilde U_{l_{12}} = \Psi_{l_{12}} - U_{l_{12}}; \quad
\Psi_{l_{12}} = D e^{-2(l_{12}- 1)/\alpha_R}, \quad
U_{l_{12}} = 2 D e^{-(l_{12}- 1)/\alpha_R}.
\label{0d4fb}
\ee
Here $\alpha_R = \alpha / R_0$, and $\alpha$ is  the effective interaction
radius. The parameter $R_0$ corresponds to the minimum of the function
$\tilde U_{l_{12}}$ ($\tilde U(l_{12}=1)=-D$ determines the depth of potential
well). It should be noted that in terms of convenience, the $R_0$-units are used
for length measuring. As a result, $R_0$- and $R_0^3$-units are used for
the linear size of each cell $c$ and volume $v$, respectively.

In the set of the collective variables $\rho_{\vk}$, the general functional
representation of the grand partition function of the cell fluid model
has the form (see \cite{kpd118,kdp117})
\bea
&&
\Xi = \int (d\rho)^{N_v} \exp \left[ \beta \mu \sqrt{N_v} \rho_{0} +
\frac{\beta}{2} \sum \limits_{\vk \in\cB} W(k) \rho_{\vk} \rho_{-\vk} \right] \non
&&
\times \prod \limits_{l=1}^{N_v} \left[ \sli_{m=0}^\infty \frac{v^m}{m!} e^{-pm^2}
\delta(\rho_{\vl}-m)\right].
\label{0d5fb}
\eea
As is seen from Eq. (\ref{0d5fb}), the occupation numbers of cells $\rho_{\vl}(\eta)$ can
take on values $m = 0, 1, 2, \ldots$. Due to the term $e^{-pm^2}$, the probability
of hosting many particles in a single cell is very small. It should be noted that
\[
(d\rho)^{N_v} = \prod \limits_{\vk\in\cB} d \rho_{\vk}.
\]
The wave vector $\vk$ belongs to the set
\[
\cB = \Big\{ \vk \!\!=\!\! (k_1, k_2, k_3)\Big| k_{i} \!\!=\!\! -\frac{\pi}{c}+
\frac{2\pi}{c}\frac{n_i}{N_a};
n_i=1,2,\ldots,N_a; i=1,2,3; N_v=N_a^3 \Big\}.
\]
The Brillouin zone $\cB$ corresponds to the volume of periodicity $\Lambda$ with cyclic
boundary conditions. The parameter $p$ characterizing the reference system is defined as
\[
p = \frac{1}{2}\beta \chi \Psi (0),
\]
and
\be
W(k) = U(k) - \Psi(k) + \chi \Psi(0)
\label{0d6fb}
\ee
is the Fourier transform of the effective potential of interaction.
The Fourier transforms of the attractive and repulsive parts of the Morse
potential ($U_{l_{12}}$ and $\Psi_{l_{12}}$ in Eqs. (\ref{0d4fb}), respectively)
satisfy the expressions
\be
U(k) = U(0) \lp 1 + \alpha_R^2 k^2 \rp^{-2}, \quad
\Psi(k) = \Psi(0) \lp 1 + \alpha_R^2 k^2 /4 \rp^{-2},
\label{0d7fb}
\ee
where
\be
\quad U(0) = 16 D \pi\frac{\alpha_R^3}{v} e^{R_0/\alpha}, \quad
\Psi(0) = D \pi \frac{\alpha_R^3}{v} e^{2R_0/\alpha}.
\label{0d8fb}
\ee
The positive parameter $\chi$ forms the Jacobian of transition to
collective variables.

The expression (\ref{0d5fb}) for the grand partition function
of the cell fluid model in the approximation of the simplest non-Gaussian
quartic fluctuation distribution (the $\rho^4$ model) can be written as
\cite{kpd118,kdp117}
\bea
&&
\Xi = g_W e^{N_v( E_\mu-a_0)} \int (d\rho)^{N_v} \exp\Biggl[ M N_v^{1/2} \rho_0 \non
&&
- \frac{1}{2} \sli_{\vk\in\cB} d(k) \rhok\rhomk \non
&&
- \frac{a_4}{4!} N_v^{-1} \sli_{{\vk_1,\ldots,\vk_4}\atop{\vk_i\in\cB}}
\rho_{\vk_1}\cdots\rho_{\vk_4} \delta_{\vk_1+\cdots+\vk_4}\Biggr].
\label{0d9fb}
\eea
Here
\bea
&&
g_W = \prod_{\vk\in\cB} \lp 2\pi \beta W(k)\rp^{-1/2},\non
&&
E_\mu = - \frac{\beta W(0)}{2} (M+\tilde a_1)^2 + M a_{34} +
\frac{1}{2} d(0) a_{34}^2 - \frac{a_4}{24} a_{34}^4,\non
&&
a_{34} = - a_3 / a_4,\non
&&
M = \mu/W(0) - \tilde a_1, \quad \tilde a_1 = a_1 + d(0) a_{34} +
\frac{a_4}{6} a_{34}^3, \non
&&
d(k) = \frac{1}{\beta W(k)}-\tilde a_2, \quad
\tilde a_2 = \frac{a_4}{2} a_{34}^2 - a_2,
\label{0d10fb}
\eea
and $\delta_{\vk_1+\cdots+\vk_4}$ is the Kronecker symbol.
In the case when $R_0/\alpha = 2.9544$ (which is typical of Na
\cite{singh}), $\chi = 1.1243$ ($p = 1.8100$), and  $v = 2.4191$
(see \cite{kpd118,kdp117}), we obtain the following quantities:
\bea
&&
a_0 = -0.3350, \quad a_1 = -0.2862, \quad a_2 = -0.2073, \non
&&
a_3 = -0.0938, \quad a_4 = 0.0376, \quad W(0) = 17.7687.\no
\eea

We shall proceed from the expression for the grand partition function
of the cell fluid model below the critical temperature
\be
\Xi = G_\mu\lp Q(r_0)\rp^{N_v} \lp \prod_{n=1}^{n'_p} Q_n\rp \Xi_{IGR}.
\label{1d1fb}
\ee
This result is obtained by step-by-step calculation of partition function within
the collective variables approach [by ``layer-by-layer'' integration in
Eq. (\ref{0d9fb}) with respect to collective variables]. The formal part of
the procedure has already been presented for $T>T_c$ in
\cite{kpd118}. Here, as in the case of $T>T_c$, we have
$G_\mu = g_W(\beta W(0))^{N_v/2} e^{N_v(E_\mu-a_0)}$.
The quantity $Q(r_0)$ appearing in Eq. (\ref{1d1fb})
corresponds to the contribution to the grand partition
function from the large values of the wave vector \cite{kpd118}, and $Q_n$ is
the partial partition function of the $n$th layer of the phase space of
the collective variables \cite{KPYu_1991}. The upper product limit $n'_p$
in formula (\ref{1d1fb}) is the number of the layer determining the point
of exit of the system from the critical regime of the order
parameter fluctuations. For $T<T_c$, the quantity $n'_p$ is defined
as \cite{KR_2009}
\be
n'_p = - \frac{\ln(\tilde h^2 + h^2_{cm})}{2\ln E_1} - 1.
\label{1d4fb}
\ee
Here the renormalized external field
\be
\tilde h = M (\beta W(0))^{1/2}
\label{1d5fb}
\ee
is a function of the chemical potential.
The temperature field
\be
h_{cm} = \tilde\tau_1^{p_0}
\label{1d6fb}
\ee
is characterized by the renormalized relative temperature
\[
\tilde\tau_1 = -\tau \frac{c_{11}}{q} E_2^{n_0}
\]
and the exponent
\[
p_0 = \frac{\ln E_1}{\ln E_2},
\]
where $\tau = (T-T_c)/T_c$, and $E_l$ are eigenvalues of the renormalization
group linear transformation matrix. The quantities $c_{11}$ and $q$ are
the same as in the case of $T>T_c$. It should be noted that in the case
of $T>T_c$ (see \cite{kpd118}), the temperature field
$h_c=\tilde\tau^{p_0}$, where $\tilde\tau=\tau c_{11}/q$, was introduced
in the expression for the exit point instead of $h_ {cm}$. The variable
$\tilde\tau$ differs from $\tilde\tau_1$ since the factor $E_2^{n_0}$
is absent. The quantity $n_0=n_p-n'_p$ (at $\tilde h=0$) characterizing
the difference between the points of exit from the critical
fluctuation regime at $T>T_c (n_p)$ and $T<T_c (n'_p)$ appears due to
a nonzero order parameter for $T<T_c$.

The quantity
\be
\Xi_{IGR} = 2^{(N_{n'_p+1}-1)/2}[Q(P_{n'_p})]^{N_{n'_{p}+1}} \Xi_{n'_p+1}
\label{1d7fb}
\ee
appearing in Eq. (\ref{1d1fb}) determines the contribution
to the grand partition function from long-wave fluctuations of the order
parameter. Here $N_{n'_{p}+1} = N_v s^{-3(n'_p+1)}$, $s$ is
the parameter of division of the phase space of the collective variables $\rhok$
into layers. As in the case of $T>T_c$, the calculations will be made for
the optimal parameter $s=s^*=3.5977$ nullifying the average value of
the coefficient in the term with the second power of the variable in
the effective density of measure at a fixed point. The quantity $Q(P_n)$ is
defined in \cite{YuKP_2001,kpd118}, and $\Xi_{n'_p+1}$ satisfies
the expression
\bea
&&
\Xi_{n'_p+1} = \int (d\rho)^{N_{n'_p+1}} \exp\Biggl[ a_1^{(n'_p+1)} \sqrt{N_{n'_p+1}} \rho_0 \non
&&
- \frac{1}{2} \sli_{\vk\in\cB_{n'_p+1}} g_{n'_p+1}(k) \rhok\rhomk \non
&&
- \frac{a_4^{(n'_p+1)}}{4!} N_{n'_p+1}^{-1} \sli_{{\vk_1,\ldots,\vk_4}\atop{\vk_i\in\cB_{n'_p+1}}}
\rho_{\vk_1}\cdots\rho_{\vk_4} \delta_{\vk_1+\cdots+\vk_4}\Biggr].
\label{1d8fb}
\eea
The coefficients $a_1^{(n)}$, $g_{n}(0)$, and $a_4^{(n)}$ are connected with the
coefficients for the ($n+1$)th layer through the recurrence relations (see \cite{kpd118}),
whose solutions are used for calculating the thermodynamic potential of the fluid system.
The quantity $b$ appearing in the equality $g_{n'_p+1}(k) = g_{n'_p+1}(0) + 2b k^2$ and
determining the coefficient in the term with the second power of the wave vector
magnitude $k$ is also given in \cite{kpd118}.

It should be noted that the above-mentioned quantity $n_0$ for a three-dimensional Ising-like
system can be found by comparing the ratio of the critical amplitudes of the correlation length
for $T>T_c$ and $T<T_c$ with data of numerical calculation from
\cite{Engels_2003}. For $s=s^*$, we obtain $n_0=0.500$ (see \cite{K_2012}).

\section{The scheme of calculating the thermodynamic potential of a model in
the tempera\-ture region below $T_c$}
\label{sec:3}

As in the case of $T>T_c$ \cite{kpd118}, we shall calculate the thermodynamic potential
$\Omega = - kT\ln\Xi$ by separating the contributions from short- and long-wave fluctuations
of the order parameter ($\Omega^{(-)}_{CR}$ and $\Omega_{IGR}$, respectively).
For $T<T_c$, we have
\be
\Omega = \Omega_\mu + \Omega_r + \Omega^{(-)}_{CR} + \Omega_{IGR}.
\label{2d1fb}
\ee
Each of the terms is the contribution of a certain multiplier of the expression (\ref{1d1fb}).
Components $\Omega_\mu$ and $\Omega_r$ satisfy the same expressions as for $T>T_c$.

The term $\Omega^{(-)}_{CR}$ in Eq. (\ref{2d1fb}) corresponds to the contribution
to the thermodynamic potential from the critical fluctuation regime. According to Eq.
(\ref{1d1fb}), we find
\be
\Omega^{(-)}_{CR} = - k T \sum_{n=1}^{n'_p} N_n f_n,
\label{2d2fb}
\ee
where $N_n = N_v s^{-3n}$. For the function $f_n(x_n, y_{n-1})$, we obtain
\be
f_n = \frac{1}{2} \ln y_{n-1} + \frac{9}{4} y^{-2}_{n-1} + \frac{x_n^2}{4} + \ln U(0, x_n).
\label{2d3fb}
\ee
Here $U(0, x_n)$ is the parabolic cylinder function \cite{Abram_1979}.
Expressions for arguments $x_n$ and $y_{n-1}$ are given in \cite{KPYu_2_1991}.
The argument $x_n$ is determined by the coefficients of the quartic fluctuation
distribution in the $n$th layer of the phase space of the collective variables.
The argument $y_{n-1}$ is the function of $x_{n-1}$.
It should be noted that taking into account Eq. (\ref{1d4fb}), we arrive at the
equalities
\bea
&&
s^{-(n'_p+1)} = \lp \tilde h^2 + h^2_{cm}\rp^{\frac{1}{d+2}}, \quad
E_1^{n'_p+1} = \lp \tilde h^2 + h^2_{cm}\rp^{-\frac{1}{2}}, \non
&&
\tilde\tau E_2^{n'_p+1} = -H_{cm}, \quad
H_{cm} = -\tilde\tau \lp \tilde h^2 + h^2_{cm}\rp^{-\frac{1}{2p_0}},\non
&&
E_{3}^{n'_p+1} = H_{3m}, \quad
H_{3m} = \lp \tilde h^2 + h^2_{cm}\rp^{\frac{\Delta}{2p_0}},
\label{2d4fb}
\eea
where $p_0$ can be represented as $p_0 = (d+2) \nu/2 $ ($d=3$ is the space dimension,
$\nu = \ln s^* / \ln E_2$ is the critical exponent of the correlation length),
$\Delta = -\ln E_3 / \ln E_2$ is the correction-to-scaling exponent.
The renormalization group parameter $s^* = 3.5977$ corresponds to the case when
the value of $x_n$ at a fixed point vanishes ($x^*=0$) \cite{K_2012}.
For the $\rho^4$ model, we have $\nu = 0.605$, $p_0 = 1.512$, $\Delta = 0.465$.
As a result of the calculation of $\Omega_{CR}^{(-)}$, we obtain
\be
\Omega_{CR}^{(-)} = - k T N_v \lp \gamma_{01} + \gamma_{02} \tau + \gamma_{03} \tau^2 \rp +
\Omega_{CR}^{(s)'}.
\label{2d5fb}
\ee
The formulas for the coefficients $\gamma_{0l}$ appearing in the analytic part of this
expression coincide with the corresponding quantities for $T>T_c$
(see \cite{kpd118}). Singular part has the form
\[
\Omega_{CR}^{(s)'} = k T N_v \bar\gamma^{(-)} s^{-3(n'_p+1)}.
\]
Here the coefficient
\be
\bar\gamma^{(-)}  = \bar\gamma_1 - \bar\gamma_2 H_{cm} + \bar\gamma_3 H_{cm}^2
\label{2d7fb}
\ee
is the function of $H_{cm}$. The constant quantities $\bar\gamma_l$ are given
in \cite{kpd118,K_2012}.

The term $\Omega_{IGR}$ in Eq. (\ref{2d1fb}) is defined as
\[
\Omega_{IGR} = - k T \ln \Xi_{IGR},
\]
where the quantity $\Xi_{IGR}$ is presented in Eq. (\ref{1d7fb}).
The contribution to the thermodynamic potential from the inverse Gaussian regime
(like the contribution to the thermodynamic potential from the limiting Gaussian
regime for $T>T_c$) can be presented in the form
\be
\Omega_{IGR} = \Omega_{TR}^{(-)} + \Omega''.
\label{2d9fb}
\ee
Here term
\be
\Omega_{TR}^{(-)} = - k T N_v f_{n'_p+1} s^{-3(n'_p+1)}
\label{2d10fb}
\ee
corresponds to the transition region from non-Gaussian to Gaussian fluctuations
of the order parameter. The term $\Omega''$ will be calculated in Sec.~IV by using the
Gaussian distribution of fluctuations.
The coefficient $f_{n'_p+1}$ corresponds to $f_n$ from Eq. (\ref{2d3fb})
for $n = n'_p+1$. It is a function of $x_{n'_p+1}$ and the quantity $y_{n'_p}$,
which in turn is expressed by $x_{n'_p}$. The expression
\be
x_{n'_p+m} = - \bar x E_2^{m-1} H_{cm} (1 - \Phi_q E_2^{m-1} H_{cm})^{-1/2}
\label{2d11fb}
\ee
can be found on the basis of solutions of the recurrence relations, taking into
account Eqs. (\ref{2d4fb}). The quantities $\bar x$ and $\Phi_q$ are the same as
in the case of $T>T_c$ (see \cite{kpd118}). It should be noted that
the terms proportional to $H_{3m}E_3^{m-1}$ are neglected when
calculating $x_{n'_p+m}$, Eq. (\ref{2d11fb}), since the quantity $H_{3m}$
from Eqs. (\ref{2d4fb}) is small near the phase transition point, and $E_3<1$.
Thus, we disregard the corrections to scaling in the present calculations.

Calculating $\Omega_{TR}^{(-)}$, Eq. (\ref{2d10fb}), we single out the transition
region from non-Gaussian to Gaussian fluctuations of the order parameter, which
corresponds to one layer of the phase space of the collective variables.
As follows from the results of \cite{YuKP_2001}, singling out
the transition region at $T<T_c$ is optional. Introducing this region
for temperatures $T<T_c$, we followed the thermodynamic potential
calculation scheme given in \cite{kpd118} for $T>T_c$.
It does not diminish the generalization of our study in any way but it does
allow us to unify the calculation scheme.

\section{Forming inverse Gaussian distribution of fluctuations}
\label{sec:4}

Let us calculate the contribution to the thermodynamic potential
$\Omega'' = - k T \ln \Xi''$ appearing in Eq. (\ref{2d9fb}).
For that we need to calculate the corresponding component of the grand
partition function
\be
\Xi'' = 2^{(N_{n'_p+2}-1)/2} [Q(P_{n'_p+1})]^{N_{n'_p+2}} \Xi_{n'_p+2},
\label{3d1fb}
\ee
where
\bea
&&
\Xi_{n'_p+2} = \int (d\rho)^{N_{n'_p+2}} \exp\Biggl[\tilde h \sqrt{N_v} \rho_0 \non
&&
- \frac{1}{2} \sli_{\vk\in\cB_{n'_p+2}} g_{n_p+2}(k) \rhok\rhomk \non
&&
- \frac{a_4^{(n'_p+2)}}{4!} N_{n'_p+2}^{-1} \sli_{{\vk_1,\ldots,\vk_4}\atop{\vk_i\in\cB_{n'_p+2}}}
\rho_{\vk_1}\cdots\rho_{\vk_4} \delta_{\vk_1+\cdots+\vk_4}\Biggr].
\label{3d2fb}
\eea
For the coefficients $g_{n'_p+2}(k)$ and $a_4^{(n'_p+2)}$, we obtain
\bea
&&
g_{n'_p+2} (k) = g_{n'_p+2} (0) + 2 b k^2,\non
&&
g_{n'_p+2} (0) = s^{-2(n'_p+2)} r_{n'_p+2},\non
&&
a_4^{(n'_p+2)} = s^{-4(n'_p+2)} u_{n'_p+2}.\no
\eea
The quantities $r_{n'_p+2}$ and $u_{n'_p+2}$ are determined by the relations
\bea
&&
r_{n'_p+2} = q (-1 - E_2 H_{cm}),\non
&&
u_{n'_p+2} = u^* (1 - \Phi_q E_2 H_{cm}).\no
\eea
Here $u^* = [q(1-s^{-2})\sqrt 3 U(x^*)[1+3/(2y^*)^2]]^2$ is the quantity
$u_n = s^{4n} a_4^{(n)}$ at a fixed point (see \cite{kpd118}).
The quantity $\tilde h$ appearing in Eq. (\ref{3d2fb}) is characterized by
the chemical potential. It corresponds to an introduction of an external
magnetic field in a three-dimensional Ising-like system
\cite{kpp406,kpp606,p806}.

\begin{figure}[htbp]
\begin{center}
\includegraphics[width=0.75\textwidth]{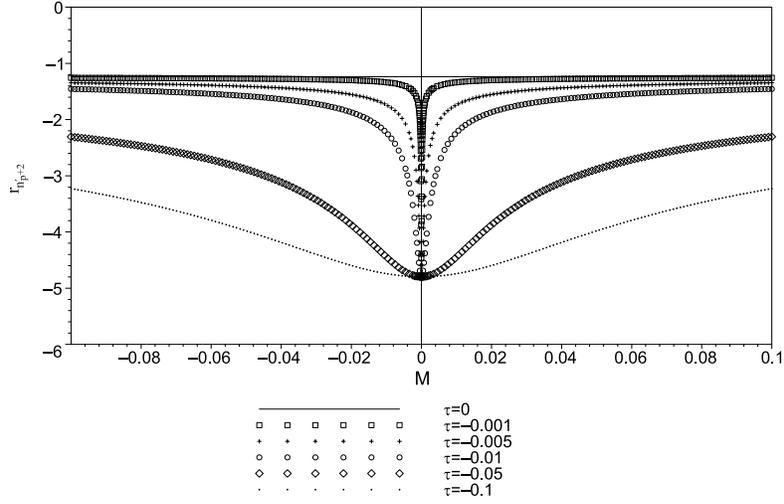}
\\
a
\\ [4ex]
\includegraphics[width=0.75\textwidth]{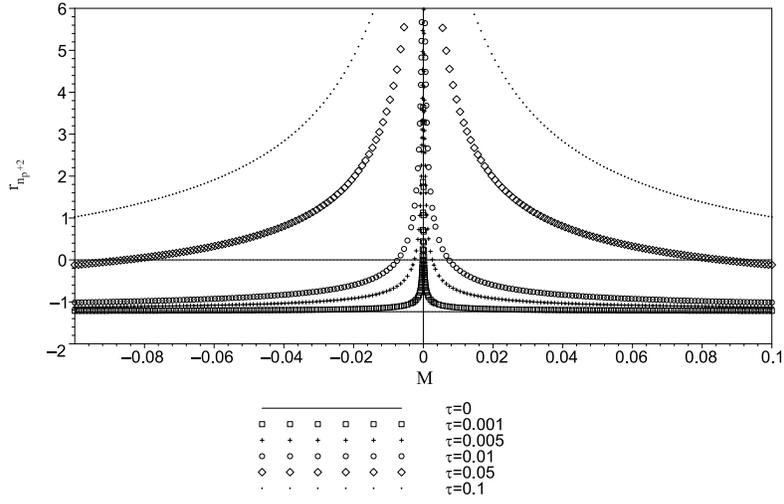}
\\
b
\end{center}
\caption{Dependence of the quantity $r_{n'_p+2}$ in the case of $T<T_c$
(figure a) and the analogous quantity $r_{n_p+2}$ for $T>T_c$ (figure b) on
the chemical potential $M$ for various values of the relative temperature $\tau$.}
\label{fig_1fb}
\end{figure}

\begin{figure}[htbp]
\centering \includegraphics[width=0.65\textwidth]{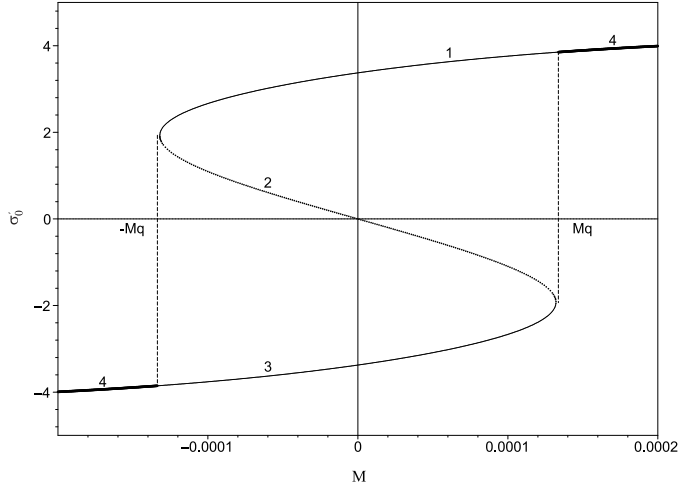}
\caption{Solutions of the cubic equation (\ref{3d12fb}) as functions
of the chemical the potential $M$ for $\tau=-0.005$. Curves 1, 2, 3, and 4
correspond to solutions $\sigma'_0=\sigma'_{01}$, $\sigma'_0=\sigma'_{02}$,
$\sigma'_0=\sigma'_{03}$, and $\sigma'_0=\sigma'_{0b}$, respectively.}
\label{fig_2fb}
\end{figure}

The coefficient $u_{n'_p+2}$ remains positive for arbitrary values of
temperature and chemical potential, so that the integral in Eq. (\ref{3d2fb}) converges.
The behavior of the coefficient $r_{n'_p+2}$, as well as of the analogous
quantity $r_{n_p+2}$ for $T>T_c$, depending on the variation of $\tau$ and $M$
is shown in Fig.~\ref{fig_1fb}.
As is seen from Fig.~\ref{fig_1fb}b, the coefficient $r_{n_p+2}$ decreases
with increasing magnitude of the chemical potential $|M|$ and becomes negative
for small values of the relative temperature $\tau$. The coefficient $r_{n'_p+2}$
takes on negative values for all temperatures $T<T_c$ (see Fig.~\ref{fig_1fb}a).
As a result, using the Gaussian approximation does not make sense when
calculating $\Xi_{n'_p+2}$, Eq. (\ref{3d2fb}). The situation can be improved by introducing
in the expression (\ref{3d2fb}) the substitution of variables
\be
\rhok = \etak + \sqrt{N_v} \sigma_- \delta_{\vk},
\label{3d5fb}
\ee
where $\sigma_-$ is a constant quantity. As a result of the substitution (\ref{3d5fb}),
the expression (\ref{3d2fb}) would take the form
\bea
&&
\Xi_{n'_p + 2} = e^{N_v E_0(\sigma_-)} \int (d\eta)^{N_{n'_p+2}} \exp \Biggl[ A'_0 \sqrt{N_v} \eta_0 \non
&&
- \frac{1}{2} \sli_{\vk\in\cB_{n'_p+2}} \bar{g'}(k) \eta_{\vk} \eta_{-\vk} -
\frac{\bar{b'}}{6} N_{n'_p+2}^{-1/2} \sli_{{\vk_{1},\ldots,\vk_{3}}\atop{\vk_i \in\cB_{n'_p+2}}}
\eta_{\vk_{1}}\cdots\eta_{\vk_{3}} \delta_{\vk_{1} + \cdots+\vk_{3}} \non
&&
- \frac{\bar{a'}_4}{24} N_{n'_p+2}^{-1} \sli_{{\vk_{1},\ldots,\vk_{4}}\atop{\vk_{i} \in\cB_{n'_p+2}}}
\eta_{\vk_{1}}\cdots\eta_{\vk_{4}} \delta_{\vk_{1} + \cdots+\vk_{4}} \Biggr].
\label{3d6fb}
\eea
Here
\be
E_0(\sigma_-) = \tilde h \sigma_- - \frac{r_{n'_p+2}}{2} s^{-2(n'_p+2)} \sigma_-^2 -
\frac{u_{n'_p+2}}{24} s^{-(n'_p+2)} \sigma_-^4.
\label{3d7fb}
\ee
For the coefficients $A'_0$, $\bar{g'}(k)$, $\bar{b'}$, and $\bar{a'}_4$, we obtain
the following expressions:
\bea
&&
A'_0 = \tilde h - r_{n'_p+2} s^{-2(n'_p+2)} \sigma_- - \frac{u_{n'_p+2}}{6} s^{-(n'_p+2)} \sigma_-^3,\non
&&
\bar{g'}(k) = \bar{g'}(0) + 2 b k^2,\non
&&
\bar{g'}(0) = r_{n'_p+2} s^{-2(n'_p+2)} + \frac{u_{n'_p+2}}{2} s^{-(n'_p+2)}\sigma_-^2,\non
&&
\bar{b'} = u_{n'_p+2} s^{-5(n'_p+2)/2} \sigma_-, \quad
\bar{a'}_4 = u_{n'_p+2} s^{-4(n'_p+2)}.
\label{3d8fb}
\eea
Similarly to the case of $T>T_c$, we find the shift quantity $\sigma_-$ from the
condition
\[
\frac{\partial E_0(\sigma_-)}{\partial \sigma_-} = 0.
\]
Taking into account Eq. (\ref{3d7fb}) and the expression for $A'_0$
[see Eqs. (\ref{3d8fb})], we obtain the equation
\be
A'_0 = 0.
\label{3d10fb}
\ee
The solution of this equation can be presented in the form
\be
\sigma_- = \sigma'_0 s^{-(n'_p+2)/2}.
\label{3d11fb}
\ee
For the quantity $\sigma'_0$, we have a cubic equation
\be
(\sigma'_0)^3 + p' \sigma'_0 + q' = 0
\label{3d12fb}
\ee
with coefficients
\[
p' = 6 \frac{r_{n'_p+2}}{u_{n'_p+2}}, \quad
q' = - 6 \frac{s^{5/2}}{u_{n'_p+2}} \frac{\tilde h}{(\tilde h^2 + h^2_{cm})^{1/2}}.
\]
In general case, the quantities $p'$ and $q'$ are functions of temperature and
chemical potential. The form of solutions of the equation (\ref{3d12fb}) depends
on the sign of the discriminant
\be
Q = (p'/3)^3 + (q'/2)^2.
\label{3d13fb}
\ee
Let us write the expression determining the magnitude of the chemical potential $M_q$
at $Q=0$. It is defined as
\be
M_q = \Biggl[ - \frac{8 r^3_{n'_p+2} (1 + \alpha^2_{mq})}{9 u_{n'_p+2} s^5 \beta W(0)}
\Biggr]^{1/2} h_{cm}.
\label{3d14fb}
\ee
Then, by substituting the quantity $\alpha_{mq} = M_q (\beta W(0))^{1/2}/h_{cm}$
in the formula (\ref{3d14fb}), we find
\[
M_q = \lb - \frac{8 r^3_{n'_p+2}}{9 u_{n'_p+2} s^5 \beta W(0)
\lp 1 + \frac{8 r^3_{n'_p+2}}{9 u_{n'_p+2} s^5} \rp}
\rb^{1/2} h_{cm}.
\]
For $Q>0$, in accordance with Cardano's formula, the real solution $\sigma'_0$
of the equation (\ref{3d12fb})  assumes the following form:
\be
\sigma'_{0b} = A+B, \quad A = (-q'/2 + Q^{1/2})^{1/3}, \quad B = (-q' / 2 - Q^{1/2})^{1/3}.
\label{3d15fb}
\ee
For $Q<0$, we have three real solutions (the quantity $\sigma'_0$
takes on three possible real values)
\bea
&&
\sigma'_{01} = 2 (-p' / 3)^{1/2} \cos (\alpha_r / 3), \non
&&
\sigma'_{02,03} = -2 (-p' / 3)^{1/2} \cos (\alpha_r / 3 \pm \pi / 3),
\label{3d16fb}
\eea
where $\alpha_r$ is determined from the relation
\[
\cos \alpha_r = - \frac{q'}{2(-p'/3)^{3/2}}.
\]
The dependence of the roots of the cubic equation (\ref{3d12fb}) on the chemical
potential $M$ for $T<T_c$ is demonstrated by the curves in Fig.~\ref{fig_2fb}.

It follows from the results of calculations that the coefficient of the quad\-ratic
term in the exponent of the expression (\ref{3d6fb}) takes on a large value in comparison
with the coefficients of other terms. We single out in Eq. (\ref{3d6fb}) the terms
with $k=0$ and perform integration with respect to variables $\eta_{\vk}$ with
$k\neq 0$ using the Gaussian distribution of fluctuations as the basis one. In a zero-order
approximation at $k\neq 0$, we arrive at the expression
\be
\Xi_{n'_p+2} = e^{N_vE_0(\sigma_-)} \prod_{{k\neq 0}\atop{\vk\in\cB_{n'_p+2}}}
\lp \pi / \bar{g'}(k) \rp^{1/2} \Xi^{(0)}_{n'_p+2}.
\label{3d18fb}
\ee
Here
\bea
&&
\Xi^{(0)}_{n'_p+2} = \int_{-\infty}^{+\infty} d \eta_0 \exp \left[ A'_0 \sqrt{N_v} \eta_0 -
\frac{1}{2} \bar{g'}(0) \eta_0^2 - \frac{\bar{b'}}{6} N_{n'_p+2}^{-1/2} \eta_0^3 \rd\non
&&
\ld - \frac{\bar{a'}_4}{24} N^{-1}_{n'_p+2} \eta_0^4 \right].
\label{3d19fb}
\eea
Eliminating the cubic term in the exponent of the expression for
$\Xi_{n'_p+2}^{(0)}$ by the substitution of the variable
\be
\eta_0 = \rho_0 - \sqrt{N_v} \sigma_-
\label{3d20fb}
\ee
and using the steepest-descent method in integrating with respect
to the variable $\rho_0$, we obtain the following formula for the
thermodynamic potential contribution corresponding
to $\Xi_{n'_p+2}$ from Eq. (\ref{3d18fb}):
\be
\Omega_{n'_p+2} = -k T N_v E_0(\sigma_-) - \frac{1}{2} k T N_{n'_p+2} \ln \pi +
\frac{1}{2} k T \sli_{{k\neq 0}\atop{\vk\in\cB_{n'_p+2}}} \ln \bar{g'}(k).
\label{3d21fb}
\ee
Here
\be
E_0(\sigma_-) = e_0^{(-)} \tilde h s^{-(n'_p+1)/2} - e_2^{(-)} s^{-3(n'_p+1)},
\label{3d22fb}
\ee
and the coefficients $e_0^{(-)}$ and $e_2^{(-)}$ satisfy the relations
\be
e_0^{(-)} = \sigma'_0 s^{-1/2}, \quad e_2^{(-)} = \frac{(\sigma'_0)^2}{2} s^{-3}
\left[ r_{n'_p+2} + \frac{u_{n'_p+2}}{12} (\sigma'_0)^2 \right].
\label{3d23fb}
\ee
It should be noted that the change of the variable $\eta_0$ [see Eq. (\ref{3d20fb})]
in the expression (\ref{3d19fb}) and the substitution $\rho_0=\sqrt{N_v}\rho$ lead to
an appearance of a sharp maximum of the integrand at the point $\bar\rho$. The extremum
condition of the integrand, from which $\bar\rho$ is determined, and the representation
$\bar\rho = \bar{\rho''}s^{-(n'_p+2)/2}$ lead to the same cubic equation for $\bar{\rho''}$
(with the same coefficients) as for $\sigma'_0$ [see Eq. (\ref{3d12fb})].
The quantities $\bar{\rho''}$ and $\sigma'_0$ as well as $\bar\rho$ and $\sigma_-$
take on the same values. This allows us to represent the integrand $E_0(\rho)$
at the point $\bar\rho$ as $E_0(\sigma_-)$ [see Eq. (\ref{3d22fb})] with
coefficients from Eqs. (\ref{3d23fb}), where the quantity $\bar{\rho''}$ is replaced
by $\sigma'_0$. Replacing summation over $\vk\in\cB_{n'_p+2}$ by integration, we arrive
at the following formula for the last term in Eq. (\ref{3d21fb}):
\bea
&&
\frac{1}{2} \sli_{\vk\in\cB_{n'_p+2}} \ln \bar{g'}(k) = k T
N_{n'_p+2} \left\{ \frac{1}{2} \ln(1+a_I^2) - (n'_p+2)\ln s \rd\non
&&
\ld + \frac{1}{2} \ln r'_R - \frac{1}{3} + \frac{1}{a_I^2} - \frac{1}{a_I^3} \arctan a_I \right\},\non
&&
a_I = \frac{\pi}{c} \lp \frac{2b}{r'_R}\rp^{1/2}, \quad
r'_R = r_{n'_p+2} + \frac{u_{n'_p+2}}{2} (\sigma'_0)^2.
\label{3d24fb}
\eea

Now, using Eq. (\ref{3d21fb}), we can write the thermodynamic potential contribution
corresponding to $\Xi''$ from Eq. (\ref{3d1fb}) as the sum of two terms
\be
\Omega'' = \Omega_0^{(-)} + \Omega'_I.
\label{3d25fb}
\ee
The component
\[
\Omega_0^{(-)} = - k T N_v E_0(\sigma_-)
\]
is related to the shift of the variable $\rho_0$, which is determined
by the quantity that is the solution of the cubic equation. For another
component $\Omega'_I$, we obtain
\[
\Omega'_I = - k T N_{n'_p+2} f_I,
\]
where
\bea
&&
f_I = \lp - \frac{1}{2} \ln 3 + 2 \ln s + \frac{1}{2} \ln u_{n'_p+1} - \ln r'_R -
\ln U(x_{n'_p+1}) \rd \non
&&
\ld - \frac{3}{4} y^{-2}_{n'_p+1}  - f''_I \rp / 2.
\label{3d28fb}
\eea
The quantity $r'_R$ is given in Eqs. (\ref{3d24fb}). The remaining quantities
in Eq. (\ref{3d28fb}) are defined as
\bea
&&
u_{n'_p+1} = u^* (1 - \Phi_q H_{cm}),\non
&&
x_{n'_p+1} = - \bar x H_{cm} (1 - \Phi_q H_{cm})^{-1/2},\non
&&
y_{n'_p+1} = s^{3/2} U(x_{n'_p+1}) \lp 3 / \varphi(x_{n'_p+1})\rp^{1/2},\non
&&
f''_I = \ln(1 + a^2_I) - \frac{2}{3} + \frac{2}{a^2_I} - \frac{2}{a^3_I} \arctan a_I.
\label{3d29fb}
\eea

Taking into account $\Omega_{TR}^{(-)}$, Eq. (\ref{2d10fb}), and $\Omega''$,
Eq. (\ref{3d25fb}), we can find the contribution to the thermodynamic
potential $\Omega_{IGR}$, Eq. (\ref{2d9fb}).

Let us add up the above-obtained contributions to the thermodynamic potential
near the critical point at $T<T_c$. According to Eq. (\ref{2d1fb}), the thermodynamic
potential of a cell fluid model is represented as the sum of several terms.
Singling out the analytic and nonanalytic parts of the thermodynamic
potential, we obtain the following complete expression equivalent to Eq. (\ref{2d1fb}):
\be
\Omega = \Omega_a + \Omega_s^{(-)} + \Omega_0^{(-)}.
\label{3d30fb}
\ee
The analytic contribution has the form
\be
\Omega_a = - k T N_v \lp \gamma_{01} - \gamma_{02} |\tau| + \gamma_{03} |\tau|^2 \rp + \Omega_{01},
\label{3d31fb}
\ee
where the formulas for $\gamma_{01}$, $\gamma_{02}$, $\gamma_{03}$, and
$\Omega_{01}= - kTN_v \left[E_\mu+\gamma_a \right]$
are the same as in the case of $T>T_c$. The nonanalytic contribution
satisfies the relation
\be
\Omega_s^{(-)} = - k T N_v \gamma_s^{(-)} \lp \tilde h^2 + h^2_{cm} \rp^{\frac{d}{d+2}}.
\label{3d32fb}
\ee
Here
\be
\gamma_s^{(-)} = f_{n'_p+1} - \bar\gamma^{(-)} + f_I / s^3.
\label{3d33fb}
\ee
The term
\bea
&&
\Omega_0^{(-)} = - k T N_v \left[ e_0^{(-)} \tilde h (\tilde h^2 + h^2_{cm})^{\frac{d-2}{2(d+2)}} \rd \non
&&
\ld - e_2^{(-)} (\tilde h^2 + h^2_{cm})^{\frac{d}{d+2}} \right],
\label{3d34fb}
\eea
is characterized by the coefficients $e_0^{(-)}$ and
$e_2^{(-)}$ given in Eqs. (\ref{3d23fb}).

Comparing the temperature regions below $T_c$ (the present calculations)
and above $T_c$ (see \cite{kpd118}), it should be noted that
the value of the system's point of exit from the critical fluctuation
regime for $T<T_c$ is somewhat smaller than the analogous value for $T>T_c$.
For this reason, we have different temperature measurement scales for each
temperature interval. Using the obtained results, we can write
the general formula for the thermodynamic potential
\be
\Omega = \Omega_a + \Omega_s + \Omega_0,
\label{3d35fb}
\ee
where the analytic part $\Omega_a$ is common to both temperature intervals,
and the contributions $\Omega_s$ and $\Omega_0$ are given by the expressions
\[
\Omega_s = \left\{
\begin{array}{l}
\Omega_s^{(-)},\quad \texttt{at} \quad T<T_c \\
\Omega_s^{(+)},\quad \texttt{at} \quad T\geq T_c
\end{array} \rd
\]
and
\[
\Omega_0 = \left\{
\begin{array}{l}
\Omega_0^{(-)},\quad \texttt{at} \quad T<T_c \\
\Omega_0^{(+)},\quad \texttt{at} \quad T\geq T_c,
\end{array} \rd
\]
respectively. Thus, we have the general functional representation of
the thermodynamic potential, which depends on both the temperature and
the chemical potential and is valid for both sides of
the critical temperature.

\section{Equation of state of the model at $T<T_c$ with allowance for
the order parameter \\ fluctuations}
\label{sec:5}

We shall use the relation
\be
\bar N = \frac{\partial \ln\Xi}{\partial \beta\mu},
\label{4d1fb}
\ee
which makes it possible to express the chemical potential in terms of the average
number of particles $\bar N$ or in terms of the average density
\be
\bar n = \frac{\bar N}{N_v} = \lp \frac{\bar N}{V}\rp v.
\label{4d2fb}
\ee
Here $v$ is the volume of the elementary cell.

On the basis of Eqs. (\ref{4d1fb}) and (\ref{4d2fb}), taking into account
Eq. (\ref{3d30fb}), we obtain
\be
\bar n = \bar n_a + n_s^{(-)} + n_0^{(-)},
\label{4d3fb}
\ee
where
\bea
&&
\bar n_a  = \frac{\partial E_\mu}{\partial\beta\mu},\non
&&
n_s^{(-)} = \frac{\partial}{\partial\beta\mu} \left[ \gamma_s^{(-)}
\lp \tilde h^2 + h^2_{cm}\rp^{\frac{d}{d+2}} \right], \non
&&
n_0^{(-)} \!\!=\!\! \frac{\partial}{\partial\beta\mu} \left[ e_0^{(-)} \tilde h\!\!
\lp \tilde h^2 + h^2_{cm}\rp^{\frac{d-2}{2(d+2)}} \!\!-
e_2^{(-)}\!\! \lp \tilde h^2 + h^2_{cm}\rp^{\frac{d}{d+2}}\right].
\label{4d4fb}
\eea

In accordance with Eq. (\ref{4d3fb}) and expressions in Appendix A,
the total contribution to the average density assumes
the following form:
\be
\bar n = - M - \tilde a_1 + \frac{1}{\beta W(0)} a_{34} +
\sigma_{00}^{(-)} \lp \tilde h^2 + h^2_{cm} \rp^{\frac{d-2}{2(d+2)}}.
\label{4d8fb}
\ee
For the coefficient $\sigma_{00}^{(-)}$, we have
\bea
&&
\sigma_{00}^{(-)} = e_0^{(-)} \frac{1}{(\beta W(0))^{1/2}}
\lp 1 + \frac{d-2}{d+2} \frac{\tilde h^2}{\tilde h^2 + h^2_{cm}} \rp \non
&&
+ e_{00}^{(-)} \frac{\tilde h}{(\tilde h^2 + h^2_{cm})^{1/2}} + e_{02}^{(-)}.
\label{4d9fb}
\eea
Here the quantity $e_0^{(-)}$ is defined in Eqs. (\ref{3d23fb}), and for
$e_{00}^{(-)}$ and $e_{02}^{(-)}$, we find expressions
\bea
&&
e_{00}^{(-)} = \frac{2d}{d+2} \frac{1}{(\beta W(0))^{1/2}} \lp \gamma_s^{(-)} - e_2^{(-)} \rp,\non
&&
e_{02}^{(-)} = \lp \frac{\partial \gamma_s^{(-)}}{\partial\beta\mu} -
\frac{\partial e^{(-)}_2}{\partial\beta\mu}\rp \lp \tilde h^2 + h^2_{cm}\rp^{1/2}.
\label{4d10fb}
\eea
The first two terms on the right-hand side of the equality (\ref{4d9fb}) depend
only on the variable $\alpha_m = \tilde h / h_{cm}$. For quantity
$\sigma_{00}^{(-)}$ to be a function of only $\alpha_m$, it is
necessary for term $e_{02}^{(-)}$ to be such a function. This can be
verified by direct differentiation of the quantities $\gamma_s^{(-)}$
and $e_2^{(-)}$ with respect to $\beta\mu$ (see Appendix B).

Taking into account the relations (\ref{0d10fb}) for $\tilde a_1$, $d(0)$, and $\tilde a_2$,
we rewrite the equation (\ref{4d8fb}) in the following form:
\be
\bar n = n_g - M + \sigma_{00}^{(-)} \lp \tilde h^2 + h^2_{cm} \rp^{\frac{d-2}{2(d+2)}}.
\label{4d34fb}
\ee
Here
\be
n_g = - a_1 - a_2 a_{34} + \frac{a_4}{3} a_{34}^3.
\label{4d35fb}
\ee
The coefficient $\sigma_{00}^{(-)}$ is given in Appendix B by
Eq. (\ref{4d32fb}). The quantity $\tilde h$ as a function of $M$ is defined
in Eq. (\ref{1d5fb}).

The nonlinear equation (\ref{4d34fb}) describes the relationship between the density
$\bar n$ and the chemical potential $M$ given in Eqs. (\ref{0d10fb}). It can be
presented in the form
\be
\bar n - n_g + M = \left( \frac{M b_1^{(-)}}{b_2^{(-)}}\right)^{1/5} \sigma_{00}^{(-)}
\label{4d36fb}
\ee
or
\be
b_3^{(-)} M^{1/5}  = \bar n - n_g + M,
\label{4d37fb}
\ee
where
\bea
&&
b_1^{(-)} = \lp \beta W(0) \rp^{1/2}, \quad
b_2^{(-)} = \frac{\alpha_m}{(1+\alpha_m^2)^{1/2}},\non
&&
b_3^{(-)} = \lp \frac{b_1^{(-)}}{b_2^{(-)}} \rp^{1/5} \sigma_{00}^{(-)}.
\label{4d38fb}
\eea

\begin{figure}[htbp]
\centering \includegraphics[width=0.80\textwidth]{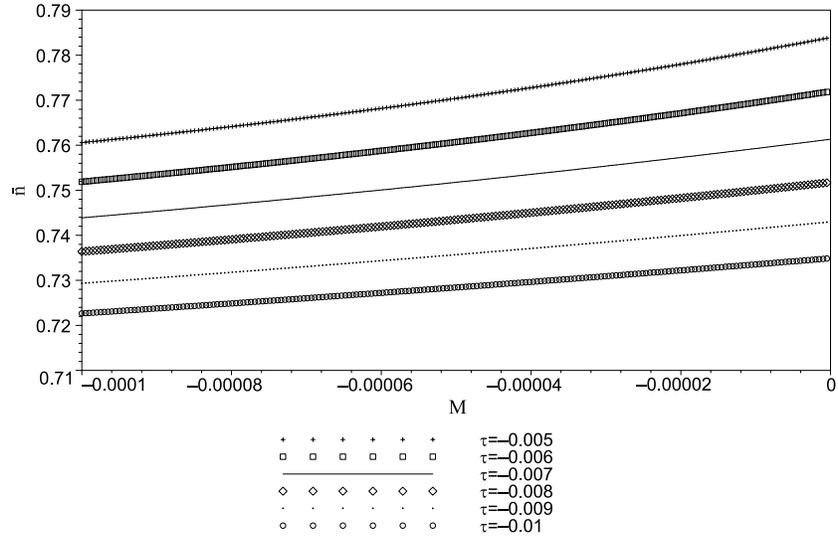}
\caption{Density $\bar n$ at negative values of the chemical
potential $M$ for some values of $\tau$.}
\label{fig_3fb}
\end{figure}

\begin{figure}[htbp]
\centering \includegraphics[width=0.80\textwidth]{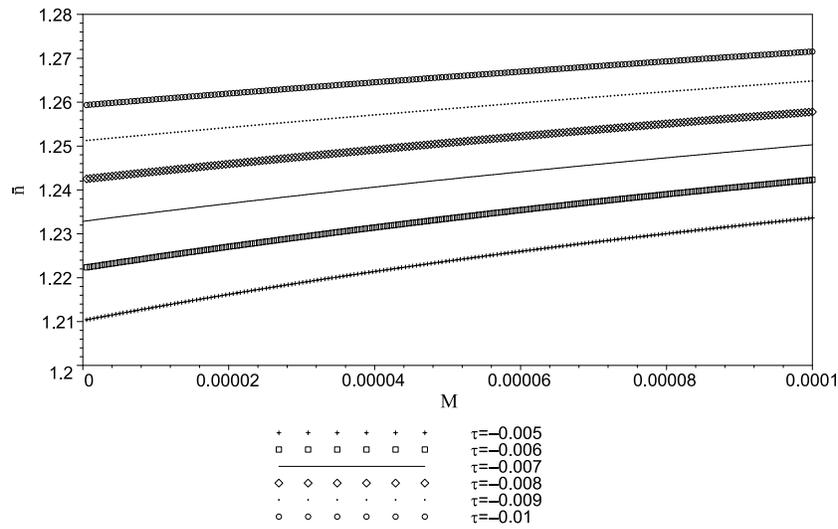}
\caption{Density $\bar n$ at positive values of the chemical
potential of $M$ for various $\tau$.}
\label{fig_4fb}
\end{figure}

Using Eq. (\ref{4d37fb}), we can express the chemical potential $M$ in terms of
the average density $\bar n$. Considering that $M\ll 1$ and neglecting the last
term on the right-hand side of the equation (\ref{4d37fb}), we obtain
an approximate formula
\be
M = \lp \frac{\bar n - n_g}{\sigma_{00}^{(-)}}\rp^5 \frac{b_2^{(-)}}{(\beta W(0))^{1/2}}
\label{4d39fb}
\ee
or (taking into account the expressions for $b_2^{(-)}$ and $\alpha_m$)
\[
M = \frac{h_{cm}}{(\beta W(0))^{1/2}}
\lb \lp \frac{\bar n - n_g}{\sigma_{00}^{(-)}}\rp^{10}
\frac{1}{h^2_{cm}} - 1 \rb^{1/2}.
\]
The equation (\ref{4d37fb}) allows us to trace the evolution of the average
density $\bar n \approx n_g + b_3^{(-)} M^{1/5}$ with the variation of $M$
for various negative values of the relative temperature $\tau$
(see Figs.~\ref{fig_3fb} and \ref{fig_4fb}).

Let us evaluate the values of critical exponents characterizing the behavior of
the density $\bar n$ as a function of temperature and chemical potential $M$
[see Eq. (\ref{4d34fb}), where the external field $\tilde h$ (which is
proportional to $M$) and the temperature field $h_ {cm}$ are given by
the expressions (\ref{1d5fb}) and (\ref{1d6fb}), respectively]. In the case
of $M=0$ and $T\neq T_c$, the equation (\ref{4d34fb}) can be rewritten as
$\bar n = \bar n_c + \sigma_{00}^{(-)}(M=0) \tilde\tau_1^{\beta}$.
Here the critical exponent $\beta=\nu/2=0.302$ is determined by the critical
exponent of the correlation length $\nu=\ln s^*/\ln E_2=0.605$, and
$\bar n_c=n_g$ is the critical density. In the case of $M\neq 0$ and $T=T_c$,
the equation (\ref{4d34fb}) can be represented as
$\bar n \approx \bar n_c + \sigma_{00}^{(-)}(T_c) \tilde h^{1/\delta}$,
where the critical exponent $\delta$ is equal to 5. Thus, the values of
the critical exponents $\nu$, $\beta$, and $\delta$ are nonclassical.

Taking into account the relations $PV=kT\ln\Xi$ and $\Omega=-kT\ln\Xi$ as well as
Eq. (\ref{3d30fb}) and $V=N_v v$, we obtain the following
the equation of state at $T<T_c$:
\bea
&&
\frac{P v}{k T} = P_a^{(-)}(T) + E_\mu +
\lp \gamma_s^{(-)} - e_2^{(-)} \rp \lp \tilde h^2 + h^2_{cm} \rp^{\frac{d}{d+2}} \non
&&
+ e_0^{(-)} \tilde h \lp \tilde h^2 + h^2_{cm} \rp^{\frac{d-2}{2(d+2)}},\non
&&
P_a^{(-)}(T) = \gamma_a + \gamma_{01} - \gamma_{02} |\tau| +
\gamma_{03} |\tau|^2,\non
&&
\gamma_a = f'_W - a_0 + \frac{1}{4} \ln \lp \frac{3}{u_0}\rp +
\frac{x^2}{4} + \ln U(0,x),\non
&&
f'_W \!\!=\!\! - \frac{1}{2} \ln \left[ 1 - (a')^2\right] + \frac{1}{3} + \frac{1}{(a')^2} -
\frac{1}{2(a')^3} \ln \left| \frac{1+a'}{1-a'}\right|.
\label{4d40fb}
\eea
Here $a' = \pi(2b)^{1/2}/c$. The quantity $M$ appearing in the expression
for $E_\mu$ [see Eqs. (\ref{0d10fb})] is given by Eq. (\ref{4d39fb}).
For $\tilde h$, taking into account Eq. (\ref{4d39fb}), we find
\be
\tilde h = \lp \frac{\bar n - n_g}{\sigma_{00}^{(-)}} \rp^5 b_2^{(-)},
\label{4d41fb}
\ee
and the quantity $h_{cm}$ is defined in Eq. (\ref{1d6fb}).
Determining the sum $\tilde h^2 + h^2_{cm}$ from Eq. (\ref{4d34fb}) and
substituting it in the expression for $Pv/(kT)$ [see Eqs. (\ref{4d40fb})],
we can rewrite the equation of state in the following form:
\bea
&&
\frac{P v}{k T} = P_a^{(-)}(T) + E_\mu +
\lp \frac{\bar n - n_g}{\sigma_{00}^{(-)}} \rp^6
\Biggl[ e_0^{(-)} \frac{\alpha_m}{(1+\alpha_m^2)^{1/2}}\non
&&
+ \gamma_s^{(-)} - e_2^{(-)} \Biggr].
\label{4d42fb}
\eea
The quantities $P_a^{(-)}(T)$, $E_\mu$, and $\sigma_{00}^{(-)}$ are
given in Eqs. (\ref{4d40fb}), (\ref{0d10fb}), and (\ref{4d32fb}),
respectively. The coefficients $e_0^{(-)}$, $e_2^{(-)}$, and
$\gamma_s^{(-)}$ are defined by the relations (\ref{3d23fb})
and (\ref{3d33fb}). The quantities
\bea
&&
a_1 = - \frac{T_1 (v, p)}{T_0 (v, p)}, \quad
a_2 = - \frac{T_2 (v, p)}{T_0 (v, p)} + a_1^2,\non
&&
a_3 = - \frac{T_3 (v, p)}{T_0 (v, p)} - a_1^3 + 3 a_1 a_2,\non
&&
a_4 = - \frac{T_4 (v, p)}{ T_0 (v, p)} + a_1^4 - 6 a_1^2 a_2 +
4 a_1 a_3 + 3 a_2^2
\label{4d43fb}
\eea
appearing in the expression for $E_\mu$ are presented in terms of special
functions
\be
T_n(v, p) = \sli_{m=0}^\infty \frac{v^m}{m!} m^n e^{-p m^2},
\label{4d44fb}
\ee
which are rapidly convergent series due to the condition $p>0$.
The parameter $p$ depends on temperature and is proportional to
the Fourier transform $\Psi(0)$ of the repulsive part of the interaction
potential for $k=0$ (see Sec.~\ref{sec:2}). Thus, the quantities $a_n$
are functions of temperature and microscopic parameters of the interaction
potential, in particular, of the ratio $R_0/\alpha$ characterizing real
substances (Morse fluids). Our numerical calculations of the quantities
$a_n$ are made for $T=T_c$. For the coefficients $\gamma_{0l}$ in
the term $P_a^{(-)}(T)$ in Eq. (\ref{4d42fb}), we have
\bea
&&
\gamma_{01} = s^{-3} \frac{f_{CR}^{(0)}}{(1-s^{-3})},\non
&&
\gamma_{02} = s^{-3} \frac{c_{11}d_1 E_2}{1-E_2 s^{-3}},\non
&&
\gamma_{03} = s^{-3} \left[ \frac{c_{12}d_1 E_2}{1-E_2 s^{-3}} +
\frac{c^2_{11} d_3 E_2^2}{1-E_2^2 s^{-3}} \right].
\label{4d45fb}
\eea
Here
\bea
&&
c_{11} = \beta_c W(0) D^{-1} \lp \tilde a_2 + 2R^{(0)} \beta_c W(0) a_4 (u^*)^{-1/2}\rp,\non
&&
c_{12} = - \beta_c W(0) D^{-1} \lp \tilde a_2 + 3R^{(0)} \beta_c W(0) a_4 (u^*)^{-1/2}\rp,
\label{4d46fb}
\eea
and $f_{CR}^{(0)}$, $d_l$ are ultimately expressed by the coordinates of a fixed
point \cite{kpd118,K_2012}. The quantities $D$ and $R^{(0)}$ are
determined by the eigenvalues and elements of the renormalization group
linear transformation matrix \cite{kpd118}. The form of
$\tilde a_2$ is given in Eqs. (\ref{0d10fb}).

The temperature $kT_c$, average density $\bar n_c$, and pressure $P_c$
at the critical point can be calculated using the corresponding equation presented
in \cite{kpd118}, Eq. (\ref{4d37fb}), and Eq. (\ref{4d42fb}), respectively.
Numerical values of $kT_c$, $\bar n_c$, and $P_c$ obtained for Na and K from
the present researches on the basis of the cell fluid model, from Monte Carlo
simulation results for the continuous system with the Morse potential in the grand
canonical ensemble \cite{singh}, and from experiment \cite{hensel}
are given in Table~\ref{tab_1fb}. The case of K corresponds to the set of
\begin{table}[htbp]
\caption{Numerical estimates of the critical-point
parameters $kT_c$, $\bar n_c$, and $P_c$ obtained for sodium and potassium from
the present theory, simulations \cite{singh}, and
experiment \cite{hensel}. Temperature $kT = kT'/D$, density
$\bar n = \rho R_0^3$ , and pressure $P = P' R_0^3/D$ are presented in the form
of reduced dimensionless units (the quantities $T'$, $\rho$, and $P'$ have
dimensional representations, for example, $[T'] = K$, $[\rho] = 1/m^3$,
and  $[P'] = Pa$).}
\label{tab_1fb}
\begin{tabular}{cccccccc}
\hline
\multicolumn{1}{c}{Research methods} & \multicolumn{3}{c}{Na} &
\multicolumn{4}{c}{K} \\
\cline {2-4}
\cline {6-8}
& \multicolumn{1}{c}{$kT_c$} & \multicolumn{1}{c}{$\bar n_c$} & \multicolumn{1}{c}{$P_c$}
& & \multicolumn{1}{c}{$kT_c$} & \multicolumn{1}{c}{$\bar n_c$} & \multicolumn{1}{c}{$P_c$} \\
\hline
\multicolumn{1}{l}{Theory (cell fluid model)} & 4.028 & 0.997 & 0.474 & & 3.304 & 0.935 & 0.408 \\
\multicolumn{1}{l}{Simulations}
& 5.874 & 1.430 & 2.159 & & 5.050 & 1.125 & 1.651 \\
\multicolumn{1}{l}{Experiment} & 3.713 & 1.215 & 0.415 & & 3.690 & 0.772 & 0.498 \\
\hline
\end{tabular}
\end{table}
parameters $R_0/\alpha = 3.0564$ (see \cite{singh}) and $\chi = 1.1981$
($p = 2.0100$), $v = 2.9402$ (see \cite{kdp117}). The set of parameters
in the case of Na is given in Sec.~\ref{sec:2}. The values of coordinates
of the critical point for Na, calculated by using the obtained relations,
coincide with the corresponding estimates obtained in \cite{kpd118}
from an analysis of temperatures above $T_c$.

\begin{figure}[htbp]
\centering \includegraphics[width=0.65\textwidth]{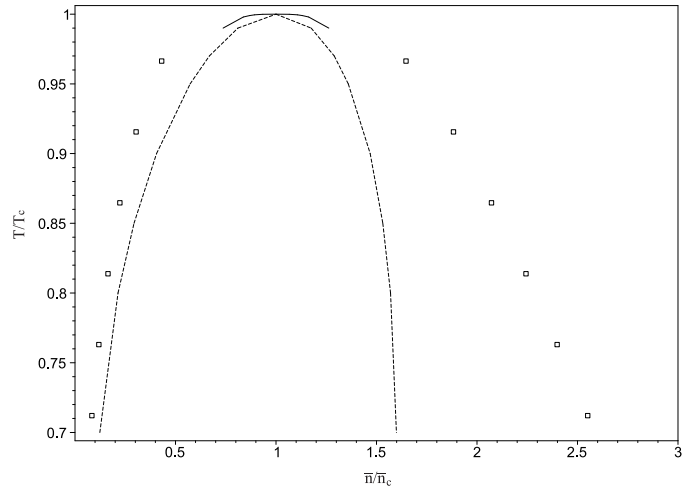}
\caption{Coexistence curve (binodal curve) for sodium. The solid curve
represents the analytic results (the present theory), the dashed curve
corresponds to results obtained in the zero mode approximation
\cite{kdp117}, and boxes represent Monte Carlo simulation
results \cite{singh}.}
\label{fig_5fb}
\end{figure}

It should be noted that the present approach is aimed at calculating and
analyzing the characteristics of the system in a narrow neighborhood of
$T_c$ ($|\tau|<\tau^*\sim 10^{-2}$, see \cite{ymo287,YuKP_2001}),
where theoretical and experimental researches are difficult to carry out.
The dome of the coexistence curve (the upper portion of the coexistence
curve) obtained for Na in the immediate vicinity of $T_c$ on the basis of
Eq. (\ref{4d34fb}) at $M=0$ is shown in Fig.~\ref{fig_5fb}
(the solid curve). For comparison of the results, the coexistence curves
obtained for Na in the case of the so called zero mode approximation
corresponding to the mean-field approximation \cite{kdp117} and
from simulation data \cite{singh} are plotted in Fig.~\ref{fig_5fb}
as well. The zero mode approximation \cite{kdp117} does not take into
account the fluctuations of the order parameter. It is not effective in
the immediate vicinity of $T_c$, where fluctuation effects play
a significant role. The approach developed in the present paper for
temperatures $T<T_c$, takes into account the fluctuation effects
that lead to the emergence of a renormalization group symmetry.
Fig.~\ref{fig_5fb} clearly demonstrates the deficiency of the
zero mode approximation to correctly obtain the branches and dome of
the binodal curve near $T_c$ (especially the liquid branch). A better
quantitative description of the dome is provided by the approach
presented here.

The description of the asymmetry of the coexistence curve will be
the subject of further studies. It is expected that the asymmetry of
the gas and liquid branches of the coexistence curve is associated with
the presence of odd powers of the variable in the Jacobian of transition
to collective variables. The description of this asymmetry in the proposed
approach is a nontrivial problem, whose solution requires a separate analysis
and additional calculations.

\section{Conclusions}
\label{sec:6}

The method for constructing the equation of state
of a cell fluid model in the high-temperature region ($T>T_c$)
(see \cite{kpd118}) is generalized to the case of $T<T_c$.
As in \cite{kpd118}, the mathematical description is performed in
the vicinity of the critical point by using the collective variables
approach with allowance for non-Gaussian fluctuations of the order parameter.
The role of the interaction potential is played by the Morse potential
possessing the Fourier transform. Calculations of the grand partition
function and thermodynamic potential of the model are carried out in
the approximation of the quartic fluctuation distribution
(the $\rho^4$ model). The complete expression for the thermodynamic
potential is obtained by summing up the contributions from all regimes
of fluctuations in accordance with a calculation technique elaborated
in this work for $T<T_c$.

For various values of the relative temperature, the variation of the fluid
density with increasing chemical potential is traced using the obtained
nonlinear equation.

The equation of state obtained in the present research in the case of
temperatures below the critical value of $T_c$ provides the pressure as
a function of temperature and density.
The main advantage of this equation of state is the presence of relations
connecting its coefficients with the fixed-point coordinates and the microscopic
parameters of the interaction potential.

This paper logically supplements the previous study for temperatures
above the critical value of $T_c$ \cite{kpd118}. As a result of this
supplement, an integral critical-behavior picture (the cases of both high-
and low-temperature regions are covered) is constructed within
the framework of a cell fluid model by using a unified approach.

Although the idea of a microscopic description of the system behavior by
the collective variables method at $T<T_c$ is similar to that used at $T>T_c$
(the separate inclusion of contributions from short-wave and long-wave
fluctuations of the order parameter), the temperature region $T<T_c$ itself
and the results obtained for it differ significantly from the case
of $T>T_c$ \cite{kpd118}. Let us point out the main differences. \\
(1) Above the critical temperature $T_c$, a liquid cannot exist.
Below the critical value of $T_c$, two phases can coexist (a gas at low
density and a liquid at high density). In contrast to $T>T_c$, the system
at $T<T_c$ acquires a nonzero order parameter (see, for example,
Fig.~\ref{fig_5fb}, where $|\bar n-\bar n_c|\neq 0$ for $T<T_c$).
It is not introduced as an independent quantity,
but is determined as a result of direct calculation. This is
possible since the set of collective variables contains
the variable $\rho_0$ associated with the order parameter. \\
(2) Due to the different quantities $n_p$ and $n'_p$,
which determine the system's points of exit from the critical fluctuation
regime at $T>T_c$ and $T<T_c$ (see Sec.~\ref{sec:2}), we have
the difference in the precise location of the critical regime
in each of the temperature cases. \\
(3) The critical regime of fluctuations is observed above
as well as below $T_c$. It is characterized by a renormalization group
symmetry and is described by a non-Gaussian distribution.
In contrast to the limiting Gaussian regime for $T>T_c$,
the inverse Gaussian regime for $T<T_c$ is described
by a non-Gaussian distribution [unlike the coefficient $r_{n_p+2}$
for $T>T_c$ (see Fig.~\ref{fig_1fb}b), the analogous quantity $r_{n'_p+2}$
appearing in the quadratic coefficient $\bar{g'}(k)$ in
an exponential function in Eq. (\ref{3d6fb}) takes on negative values
for all temperatures $T<T_c$ (see Fig.~\ref{fig_1fb}a)]. The distribution
acquires the Gaussian form [the quantity $\bar{g'}(k)$ in
Eq. (\ref{3d6fb}) becomes positive] only as a result
of the substitution (\ref{3d5fb}). \\
(4) The contribution $\Omega_0^{(-)}$ [see Eq. (\ref{3d34fb})] to
the total thermodynamic potential is characterized by the coefficients
$e_0^{(-)}$ and $e_2^{(-)}$ [see Eqs. (\ref{3d23fb})] and is
associated with the shift of the variable $\rho_0$. The cubic equation
(\ref{3d12fb}) for the quantity $\sigma'_0$ determining this shift
$\sigma_- = \sigma'_0 s^{-(n'_p+2)/2}$ is obtained. For $T>T_c$,
the analogous cubic equation has only one real root \cite{kpd118},
which is zero in the absence of the chemical potential.
The situation for $\sigma'_0$ in the temperature region $T<T_c$ is more
complicated. Unlike the case of $T>T_c$, the equation (\ref{3d12fb})
can have more than one real root. The number of real solutions depends
on the value of the chemical potential $M$. For all $|M|>M_q$
(discriminant $Q>0$), one real root exists [see Eqs. (\ref{3d15fb})].
For $|M|<M_q$ ($Q<0$), the cubic equation (\ref{3d12fb}) has three real
solutions [see Eqs. (\ref{3d16fb}) and Fig.~\ref{fig_2fb}]. The magnitude
of the chemical potential $M_q$ [see Eq. (\ref{3d14fb})] is found from
the condition $Q=0$. It should be noted that the contribution
$\Omega_0^{(-)}$ is nonzero even when $M=0$. For $T>T_c$, the analogous
contribution $\Omega_0^{(+)}$ \cite{kpd118} is equal to zero
in the case when $M=0$. \\
(5) The function $E_0(\sigma_-)$ [see Eqs. (\ref{3d7fb}) or
(\ref{3d22fb}) and (\ref{2d4fb}) for $s^{-(n'_p+1)}$] appearing in
the contribution $\Omega_0^{(-)}=-kTN_vE_0(\sigma_-)$ from the macroscopic
part of the variable $\rho_0$ coincides in form with the integrand
$E_0(\rho)$ at the extremum point $\bar\rho$, which is the same as
$\sigma_-$ (see Sec.~\ref{sec:4}). The quantity $\sigma'_0$ in
the expression (\ref{3d11fb}) for $\sigma_-$ is defined by three real
solutions of the cubic equation (\ref{3d12fb}) when the discriminant $Q<0$.
From the three real solutions (\ref{3d16fb}), we choose
one solution that leads to a maximum of $E_0(\sigma_-)$.
The solution $\sigma'_0=\sigma'_{01}$ corresponds to the maximum
of $E_0(\sigma_-)$ at the chemical potential $M>0$, and the solution
$\sigma'_0=\sigma'_{03}$ corresponds to the maximum of $E_0(\sigma_-)$
at the chemical potential $M<0$. In contrast to $T>T_c$, the order
parameter distribution characterized by the function $E_0(\sigma_-)$
exhibits a bimodal shape \cite{bl184,l196,cg103}. \\
(6) On the basis of the obtained equation (\ref{4d37fb}), the curves
describing the dependence of the density $\bar n$ on the chemical
potential $M$ are plotted for various values of temperatures $T<T_c$
(see Figs.~\ref{fig_3fb} and \ref{fig_4fb}). As is clearly seen from
Figs.~\ref{fig_3fb} and \ref{fig_4fb}, a change in sign of $M$ leads
to a jump in $\bar n$. For $T>T_c$, a similar density jump is
not observed since the cubic equation has a single solution,
which is a continuous function of $M$ (see \cite{kpd118}).

The region in the vicinity of the critical point is interesting
(due to the fundamental and applied aspects) and  difficult
(due to the essential role of fluctuation effects) to analyze.
The Morse fluid theory in the vicinity of the critical point
is constructed in this work without involving the hard-spheres reference system.
The reference system is formed as some part of the repulsive component of
the interaction potential in the course of calculating the grand partition function.
This allows us to take into account both short-range and long-range interactions
from the unified positions of the collective variables approach.

It is expected that the method developed here and in \cite{kpd118}
for Morse fluids can be applied to the description of a phase transition in simple
liquid alkali metals. Numerical estimates of the critical-point parameters for
potassium are obtained in addition to the estimates for sodium. The values of
dimesionless critical temperature, density, and pressure, calculated for Na and K
on the basis of the proposed method, are in accord with the other authors' data.
As is seen from Table~\ref{tab_1fb}, our values of the critical temperature,
density, and pressure agree more closely with experimental data than
Monte Carlo simulation results from \cite{singh}. In \cite{singh},
the authors note that the critical pressures for the alkali metals (sodium and
potassium) show large deviations from the experimental values.
The pressure for Na is vastly overestimated because
the critical temperature is overvalued (see Table~\ref{tab_1fb}). It is observed
that at the experimental critical point of Na metal, 2485 K, the corresponding
pressure predicted by simulation is a good approximation to the experimental
critical pressure. Our estimates for the critical temperature $kT_c$
in the $\rho^4$ model approximation (4.028 for Na and 3.304 for K,
see Table~\ref{tab_1fb}) also agree more closely with the experimental values
(3.713 for Na and 3.690 for K, see Table~\ref{tab_1fb}) than
the corresponding results (5.760 for Na and 5.037 for K) obtained in the previous
study \cite{kdp117} in the zero mode approximation (the mean-field
approximation). Compared with the previous results
(the dashed curve in Fig.~\ref{fig_5fb}), our estimates for the binodal dome
in the immediate vicinity of $T_c$ (the solid curve in Fig.~\ref{fig_5fb})
are in better agreement with the predicted data that can be obtained for Na
by extrapolation of the Monte Carlo simulation results
\cite{singh} (boxes in Fig.~\ref{fig_5fb}) to $T/T_c \approx 1$.

We hope that the proposed method and explicit representations obtained for a simple
fluid system may provide useful benchmarks in studying the critical behavior of
a multicomponent fluid. The performed researches also deepen knowledge about
the critical properties of fluids and serve as a certain methodological contribution
to the theoretical description of critical phenomena.

\renewcommand{\theequation}{A.\arabic{equation}}
\section*{Appendix A\\
Average density components}
\setcounter{equation}{0}
\label{appA}

Let us write the terms appearing in the expression for $\bar n$ [see
Eqs. (\ref{4d3fb}) and (\ref{4d4fb})].

The first term in Eq. (\ref{4d3fb})
\be
\bar n_a = - M - \tilde a_1 + \frac{a_{34}}{\beta W(0)}
\label{4d5fb}
\ee
corresponds to the analytic part.

The second term in Eq. (\ref{4d3fb}) is represented as
\bea
&&
n_s^{(-)} = \lp \tilde h^2 + h^2_{cm}\rp^{\frac{d-2}{2(d+2)}}
\Biggl[ \lp \tilde h^2 + h^2_{cm}\rp^{1/2}
\frac{\partial \gamma_s^{(-)}}{\partial\beta\mu} + \frac{2d}{d+2} \gamma_s^{(-)} \non
&&
\times \frac{1}{(\beta W(0))^{1/2}}
\frac{\tilde h}{\lp \tilde h^2 + h^2_{cm}\rp^{1/2}} \Biggr].
\label{4d6fb}
\eea

The third term $n_0^{(-)}$ in Eq. (\ref{4d3fb})
satisfies the relation
\bea
&&
n_0^{(-)} = \lp \tilde h^2 + h^2_{cm}\rp^{\frac{d-2}{2(d+2)}}\!\!
\Biggl[ e_0^{(-)} \frac{1}{(\beta W(0))^{1/2}}
\lp 1 + \frac{d-2}{d+2} \frac{\tilde h^2}{\tilde h^2 + h^2_{cm}} \rp\!\! \non
&&
- \frac{2d}{d+2} e_2^{(-)} \frac{1}{(\beta W(0))^{1/2}} \frac{\tilde h}{(\tilde h^2 + h^2_{cm})^{1/2}} \non
&&
- \lp \tilde h^2 + h^2_{cm}\rp^{1/2} \frac{\partial e_2^{(-)}}{\partial\beta\mu} \Biggr].
\label{4d7fb}
\eea

\renewcommand{\theequation}{B.\arabic{equation}}
\section*{Appendix B\\
Quantity $\sigma_{00}^{(-)}$ as function of $\alpha_m$}
\setcounter{equation}{0}
\label{appB}

Let us consider the derivatives $\frac{\partial e^{(-)}_2}{\partial\beta\mu}$
and $\frac{\partial \gamma_s^{(-)}}{\partial\beta\mu}$ appearing in
the expression for $e_{02}^{(-)}$ [see Eqs. (\ref{4d10fb}),
(\ref{3d23fb}), and (\ref{3d33fb})]. The final formula for $\sigma_{00}^{(-)}$
will be given later.

For the derivative of $e_2^{(-)}$ with respect to $\beta\mu$, we obtain
\be
\frac{\partial e^{(-)}_2}{\partial\beta\mu} = - \frac{1}{(\beta W(0))^{1/2}} q'_s (\sigma'_0)^2
\left[ 1 + \frac{q_l}{12} (\sigma'_0)^2 \right] (\tilde h^2 + h^2_{cm})^{-1/2},
\label{4d11fb}
\ee
where
\bea
&&
q'_s = - \frac{E_2}{2p_0} q s^{-3} H_{cm} \frac{\alpha_m}{(1+\alpha_m^2)^{1/2}},\non
&&
q_l = \Phi_q u^* q^{-1}.\no
\eea
Here, it is taken into account that
\bea
&&
\frac{\partial r_{n'_p+2}}{\partial\beta\mu} = \frac{1}{(\beta W(0))^{1/2}} q E_2 H_{cmd}
\lp \tilde h^2 + h_{cm}^2 \rp^{-1/2},\non
&&
\frac{\partial u_{n'_p+2}}{\partial\beta\mu} = \frac{1}{(\beta W(0))^{1/2}} u^* \Phi_q E_2 H_{cmd}
\lp \tilde h^2 + h_{cm}^2 \rp^{-1/2},\non
&&
\frac{\partial H_{cm}}{\partial\beta\mu} =
\frac{\partial H_{cm}}{\partial \tilde h} \frac{\partial \tilde h}{\partial\beta\mu} =
- \frac{1}{(\beta W(0))^{1/2}} H_{cmd} \lp \tilde h^2 + h^2_{cm}\rp^{-1/2},\non
&&
H_{cmd} = \frac{H_{cm}}{p_0} \frac{\alpha_m}{(1+\alpha^2_m)^{1/2}}.\no
\eea

The derivative of $\gamma_s^{(-)}$ is calculated on the basis of Eq. (\ref{3d33fb}),
where an explicit expression for each term is known. The expression
for the derivative
\be
\frac{\partial \gamma_s^{(-)}}{\partial\beta\mu} = \frac{\partial f_{n'_p+1}}{\partial\beta\mu} -
\frac{\partial \bar\gamma^{(-)}}{\partial\beta\mu} + s^{-3} \frac{\partial f_I}{\partial\beta\mu}
\label{4d14fb}
\ee
is the sum of the derivatives of the quantities $\bar\gamma^{(-)}$, $f_{n'_p+1}$,
and $f_I/s^3$, which describe contributions from the region of the critical regime
of the order parameter fluctuations, the transition region, and the region of
the inverse Gaussian regime, respectively.

Let us calculate the derivative of the quantity $f_{n'_p+1}$ [see Eq. (\ref{2d3fb})
for $n= n'_p+1$] with respect to the chemical potential. Taking into account
the relations
\bea
&&
\frac{\partial y_{n'_p+m}}{\partial\beta\mu} = y_{n'_p+m} r'_{p+m}
\frac{\partial x_{n'_p+m}}{\partial\beta\mu},\non
&&
\frac{\partial x_{n'_p+m}}{\partial\beta\mu} =  \frac{1}{(\beta W(0))^{1/2}}
g'_{p+m} \lp \tilde h^2 + h^2_{cm} \rp^{-1/2},\non
&&
U'(0, x_{n'_p+m}) \!=\! - \frac{x_{n'_p+m}}{2} U(0, x_{n'_p+m}) -
\frac{1}{2} U(1, x'_{n_p+m}),\no
\eea
where
\bea
&&
r'_{p+m} = \frac{U'(x_{n'_p+m})}{U(x_{n'_p+m})} -
\frac{1}{2} \frac{\varphi'(x_{n'_p+m})}{\varphi(x_{n'_p+m})}, \non
&&
g'_{p+m} = \bar x H_{cmd} E_2^{m-1} \lp 1 - \Phi_q H_{cm} E_2^{m-1} \rp^{-1/2} \non
&&
\times \left[ 1 + \frac{\Phi_q}{2} H_{cm} E_2^{m-1} \lp 1 - \Phi_q H_{cm} E_2^{m-1}\rp^{-1} \right],\non
&&
U'(x_{n'_p+m}) = \frac{1}{2} U^2(x_{n'_p+m})+ x_{n'_p+m} U(x_{n'_p+m}) - 1, \non
&&
\varphi'(x_{n'_p+m}) = 6 U'(x_{n'_p+m}) U(x_{n'_p+m}) + 2 U(x_{n'_p+m}) \non
&&
+ 2 x_{n'_p+m} U'(x_{n'_p+m}),\no
\eea
we find
\be
\frac{\partial f_{n'_p+1}}{\partial\beta\mu} = \frac{1}{(\beta W(0))^{1/2}}
f'_p \lp \tilde h^2 + h^2_{cm} \rp^{-1/2}.
\label{4d17fb}
\ee
Here
\[
f'_p = \frac{1}{2} r'_p g'_p \lp 1 - 9 / y^2_{n'_p}\rp - \frac{1}{2} g'_{p+1} U(x_{n'_p+1}).
\]
The derivative of the quantity $\bar\gamma^{(-)}$ [see Eq. (\ref{2d7fb})] with respect to
the chemical potential is written in the form
\be
\frac{\partial \gamma^{(-)}}{\partial\beta\mu} = - \frac{1}{(\beta W(0))^{1/2}}
\gamma'_p \lp \tilde h^2 + h^2_{cm} \rp^{-1/2},
\label{4d19fb}
\ee
where
\[
\gamma'_p = H_{cmd} \lp - \bar\gamma_2 + 2 \bar\gamma_3 H_{cm} \rp.
\]
The calculation of the derivative of $ f_I $ [see Eq. (\ref{3d28fb})] with respect to
$\beta\mu$ will be carried out taking into account that the quantity
$\sigma'_0$ is a function of the chemical potential. Using the equality (\ref{3d10fb}),
we obtain the relation
\[
\frac{\partial \sigma'_0}{\partial\beta\mu} = \frac{1}{(\beta W(0))^{1/2}}
g'_\sigma \lp \tilde h^2 + h^2_{cm} \rp^{-1/2},
\]
which makes it possible to find the following expression for the derivative of the quantity
$r'_R$ appearing in Eq. (\ref{3d28fb}):
\be
\frac{\partial r'_R}{\partial\beta\mu} = \frac{1}{(\beta W(0))^{1/2}}
g'_R \lp \tilde h^2 + h^2_{cm} \rp^{-1/2}.
\label{4d22fb}
\ee
Here
\bea
&&
g'_\sigma = \frac{s^{5/2}}{r'_R} \frac{1}{1+\alpha^2_m} -
\frac{\sigma'_0}{r'_R} H_{cmd} q E_2 \lp 1 + \frac{q_l}{6} (\sigma'_0)^2\rp,\non
&&
g'_R = q E_2 H_{cmd} \lp 1 + \frac{q_l}{2} (\sigma'_0)^2\rp + u_{n'_p+2} g'_\sigma \sigma'_0.\no
\eea
Another obtained formula
\[
\frac{\partial a_I}{\partial\beta\mu} = \frac{1}{(\beta W(0))^{1/2}}
g_{aI} \lp \tilde h^2 + h^2_{cm} \rp^{-1/2},
\]
where
\[
g_{aI} = - \frac{a_I g'_R}{2r'_R},
\]
allows us to calculate the derivative of the quantity $f''_I$
[see Eqs. (\ref{3d29fb})] appearing in the expression (\ref{3d28fb}) for $f_I$.
We have
\be
\frac{\partial f''_I}{\partial\beta\mu} = \frac{1}{(\beta W(0))^{1/2}}
g_{aI} a_{Ig} \lp \tilde h^2 + h^2_{cm} \rp^{-1/2}.
\label{4d26fb}
\ee
Here
\[
a_{Ig} = \frac{2a_I}{1+a_I^2} - \frac{4}{a_I^3} + \frac{6}{a_I^4} \arctan a_I -
\frac{2}{a_I^3} \frac{1}{1+a_I^2}.
\]
Using Eqs. (\ref{4d22fb}) and (\ref{4d26fb}), we arrive at the expression
\be
\frac{\partial f_I}{\partial\beta\mu} = \frac{1}{(\beta W(0))^{1/2}}
f_{Iv} \lp \tilde h^2 + h^2_{cm}\rp^{-1/2},
\label{4d28fb}
\ee
where
\bea
&&
f_{Iv} = \frac{1}{4} \frac{u^* \Phi_q}{u_{n'_p+1}} H_{cmd} -
\frac{1}{2} \lp \frac{g'_R}{r'_R} + g_{aI} a_{Ig}\rp \non
&&
+ g'_{p+1} \lp \frac{3}{4} \frac{r'_{p+1}}{y^2_{n'_p+1}} -
\frac{1}{2} \frac{U'(x_{n'_p+1})}{U(x_{n'_p+1})}\rp.\no
\eea

With allowance for Eqs. (\ref{4d17fb}), (\ref{4d19fb}), and (\ref{4d28fb}),
the derivative from Eq. (\ref{4d14fb}) can be represented as
\be
\frac{\partial \gamma_s^{(-)}}{\partial\beta\mu} = \frac{1}{(\beta W(0))^{1/2}}
f_{\delta_1} \lp \tilde h^2 + h^2_{cm} \rp^{-1/2}.
\label{4d30fb}
\ee
Here
\be
f_{\delta_1} = f'_p + \gamma'_p + f_{Iv} / s^3.
\label{4d31fb}
\ee
The quantities $f'_p$, $\gamma'_p$, and $f_{Iv}$  depend only
on the parameter $\alpha_m$.

The above expressions (\ref{4d10fb}), (\ref{4d11fb}), and (\ref{4d30fb}) make
it possible to write the coefficient $\sigma_{00}^{(-)}$, Eq. (\ref{4d9fb}),
in the final explicit form. We have
\bea
&&
\sigma_{00}^{(-)} = e_0^{(-)} \frac{1}{(\beta W(0))^{1/2}}
\lp 1 + \frac{d-2}{d+2} \frac{\alpha_m^2}{1+\alpha_m^2} \rp \non
&&
+ e_{00}^{(-)} \frac{\alpha_m}{(1+\alpha_m^2)^{1/2}} + e_{02}^{(-)},
\label{4d32fb}
\eea
where
\be
e_{02}^{(-)} = \frac{1}{(\beta W(0))^{1/2}}
\left[ f_{\delta_1} + q'_s (\sigma'_0)^2 \lp 1 + q_l (\sigma'_0)^2 / 12\rp \right].
\label{4d33fb}
\ee
Thus, the quantity $\sigma_{00}^{(-)}$ is a function of
\[
\alpha_m = \frac{\tilde h}{h_{cm}} = (\beta W(0))^{1/2}
\lp \frac{q}{c_{11}} E_2^{-n_0}\rp^{p_0} \alpha_0.
\]
The multiplier
\[
\alpha_0 = \frac{M}{|\tau|^{p_0}} = \frac{\mu / W(0) - \tilde a_1}{|\tau|^{p_0}}
\]
includes the initial quantities $\mu$ and $\tau$.


\begin{thebibliography}{29}

\bibitem{hm113} J.-P. Hansen, I. R. McDonald, Theory of Simple Liquids:
With Applications to Soft Matter, Academic Press, Oxford, 2013.
%
\bibitem{km189} K. N. Khanna, I. L. McLaughlin, J. Phys.: Condens.
Matter 1 (1989) 4155.
%
\bibitem{bs104} D. Ben-Amotz, G. Stell, J. Phys. Chem. B 108 (2004) 6877.
%
\bibitem{r110} R. Roth, J. Phys.: Condens. Matter 22 (2010) 063102.
%
\bibitem{y114} I. R. Yukhnovskii, Condens. Matter Phys. 17 (2014) 43001.
%
\bibitem{ymo287} I. R. Yukhnovskii, Phase Transitions of the
Second Order. Collective Variables Method, World Scientific, Singapore,
1987.
%
\bibitem{YuKP_2001} I. R. Yukhnovskii, M. P. Kozlovskii, I. V. Pylyuk,
Microscopic Theory of Phase Transitions in the Three-Dimensional Systems,
Eurosvit, Lviv, 2001 [in Ukrainian].
%
\bibitem{rebenko_13}  A. L. Rebenko, Rev. Math. Phys. 25 (2013) 1330006.
%
\bibitem{rebenko_15} V. A. Boluh, A. L. Rebenko, J. Mod. Phys. 6
(2015) 168.
%
\bibitem{okumura_00} H. Okumura, F. Yonezawa, J. Chem. Phys. 113
(2000) 9162.
%
\bibitem{singh} J. K. Singh, J. Adhikari, S. K. Kwak, Fluid Phase Equilib.
248 (2006) 1.
%
\bibitem{apf_11} E. M. Apfelbaum, J. Chem. Phys. 134 (2011) 194506.
%
\bibitem{martinez} A. Mart\'{i}nez-Valencia, M. Gonz\'{a}lez-Melchor,
P. Orea, J. L\'{o}pez-Lemus, Mol. Simul. 39 (2013) 64.
%
\bibitem{kpd118}  M. P. Kozlovskii, I. V. Pylyuk, O. A. Dobush,
Condens. Matter Phys. 21 (2018) 43502.
%
\bibitem{ykp202} I. R. Yukhnovskii, M. P. Kozlovskii, I. V. Pylyuk,
Phys. Rev. B 66 (2002) 134410.
%
\bibitem{kdp117}   M. P. Kozlovskii, O. A. Dobush, I. V. Pylyuk,
Ukr. J. Phys. 62 (2017) 865.
%
\bibitem{KPYu_1991} M. P. Kozlovskii, I. V. Pylyuk, I. R. Yukhnovskii,
Theor. Math. Phys. 87 (1991) 540.
%
\bibitem{KR_2009} M. P. Kozlovskii, R. V. Romanik, J. Phys. Stud. 13
(2009) 4007.
%
\bibitem{Engels_2003}  J. Engels,  L. Fromme, M. Seniuch, Nucl. Phys. B
655 (2003) 277.
%
\bibitem{K_2012} M. P. Kozlovskii, The Influence of an External Field on
the Critical Behavior of the Three-Dimensional Systems, Galytskii Drukar,
Lviv, 2012 [in Ukrainian].
%
\bibitem{Abram_1979} Handbook of Mathematical Functions with Formulas,
Graphs and Mathematical Tables, edited by Milton Abramowitz,
Irene A. Stegun, National Bureau of Standards, Applied Mathematics
Series 55, 1964.
%
\bibitem{KPYu_2_1991} M. P. Kozlovskii, I. V. Pylyuk, I. R. Yukhnovskii,
Theor. Math. Phys. 87 (1991) 641.
%
\bibitem{kpp406} M. P. Kozlovskii, I. V. Pylyuk, O. O. Prytula,
Phys. Rev. B 73 (2006) 174406.
%
\bibitem{kpp606} M. P. Kozlovskii, I. V. Pylyuk, O. O. Prytula,
Nucl. Phys. B 753 (2006) 242.
%
\bibitem{p806} I. V. Pylyuk, J. Magn. Magn. Mater. 305 (2006) 216.
%
\bibitem{hensel} F. Hensel, J. Phys.: Condens. Matter 2 (1990) SA33.
%
\bibitem{bl184} K. Binder, D. P. Landau, Phys. Rev. B 30
(1984) 1477.
%
\bibitem{l196} K.-C. Lee, Phys. Rev. E 53 (1996) 6558.
%
\bibitem{cg103} Ph. Chomaz, F. Gulminelli, Physica A 330
(2003) 451.
\end{thebibliography}
\end{document}